\documentclass[11pt,a4paper]{article}
\usepackage[utf8]{inputenc}
\usepackage[T1]{fontenc}
\usepackage[english]{babel}
\usepackage[margin=25mm]{geometry}
\usepackage{amsmath,amsthm}
\usepackage{newtxtext,newtxmath}
\usepackage{bm}
\usepackage{graphicx}
\usepackage{xcolor}
\usepackage{microtype}
\usepackage{titlesec}
\usepackage[font=small,labelfont=bf,labelsep=period,
  justification=justified,singlelinecheck=false]{caption}
\usepackage{placeins}
\usepackage[numbers,sort&compress]{natbib}
\usepackage[
  colorlinks=true,
  citecolor=blue,
  linkcolor=blue,
  urlcolor=blue
]{hyperref}

\hypersetup{
  pdftitle={The active chiral random walk},
  pdfauthor={Nir Sherf, Joshua Uhlig, Jan Wojcik, Ralf Metzler, Yael Roichman, Erik Kalz},
  pdfkeywords={chiral random walks, active matter, lattice models, persistent motion, odd diffusion}
}
\titleformat{\section}{\large\bfseries}{\thesection}{0.65em}{}
\titleformat{\subsection}{\normalsize\bfseries}{\thesubsection}{0.65em}{}
\titlespacing*{\section}{0pt}{2.4ex plus 0.6ex minus 0.3ex}{1ex}
\titlespacing*{\subsection}{0pt}{1.9ex plus 0.5ex minus 0.2ex}{0.7ex}
\numberwithin{equation}{section}
\newcommand{\ket}[1]{|{#1})}
\newcommand{\bra}[1]{({#1}|}
\newcommand{\braket}[2]{({#1}|{#2})}

\begin{document}

\begin{center}
  {\LARGE\bfseries The active chiral random walk\par}
  \vspace{0.8em}
  {\normalsize
  Nir Sherf$^{1}$, Joshua Uhlig$^{2}$, Jan W{\'o}jcik$^{3}$,\\[3pt]
  Ralf Metzler$^{2}$, Yael Roichman$^{1,4,5}$ and Erik Kalz$^{2,6}$\par}
  \vspace{0.65em}
  \begin{minipage}{0.97\textwidth}
    \centering\footnotesize
    $^{1}$Raymond \& Beverly Sackler School of Chemistry, Tel Aviv University, Tel Aviv 6997801, Israel\par
    \smallskip
    $^{2}$University of Potsdam, Institute of Physics and Astronomy, D-14476 Potsdam, Germany\par
    \smallskip
    $^{3}$Institute of Theoretical Physics and Astrophysics, University of Gda{\'n}sk, 80-308 Gda{\'n}sk, Poland\par
    \smallskip
    $^{4}$Raymond \& Beverly Sackler School of Physics and Astronomy, Tel Aviv University, Tel Aviv 6997801, Israel\par
    \smallskip
    $^{5}$The Center for Physics and Chemistry of Living Systems, Tel Aviv University, Tel Aviv 6997801, Israel\par
    \smallskip
    $^{6}$Institut de Sciences et d'Ing{\'e}nierie Supramol{\'e}culaires, University of Strasbourg, FR-67000 France\par
  \end{minipage}
  \par\vspace{0.55em}
  {\footnotesize Corresponding authors: Nir Sherf and Erik Kalz\\[2pt]
  \href{mailto:nirs@mail.tau.ac.il}{nirs@mail.tau.ac.il},
  \href{mailto:erikkalz98@gmail.com}{erikkalz98@gmail.com}\par}
\end{center}
\vspace{-0.4em}

\begin{abstract}
We introduce a minimal and intuitive lattice model for an active chiral particle which exhibits persistence, reversal, and chirality in its dynamics.
It is fully defined by a small set of microscopic parameters that provide transparent control over the resulting dynamics.
By coarse-graining, we extract continuum equations and closed-form expressions for the velocity autocorrelation, off-diagonal directional correlation, mean-squared displacement, and long-time diffusion and odd diffusion coefficients.
The complex parameter $\lambda=q_p+\mathrm{i}q_c$ thereby encodes the microscopic information on persistence ($q_p$) and chirality ($q_c$) of the walker.
In a weak-turning limit, microscopically realized by infrequent reorientation events, the continuum description can be mapped onto well-established models of active matter.
Outside this limit the discrete model exhibits finite-step dynamics, such as anti-persistent motion, that are not captured by a smooth angular-diffusion description.
The model therefore provides a compact, exactly solvable, and readily tunable framework in which a small number of transparent microscopic parameters control chiral active transport.
\end{abstract}

{\small\noindent\textbf{Keywords:} chiral random walks, active matter, lattice models, persistent motion, odd diffusion.\par}

\section{Introduction}

Random walks are among the simplest models of stochastic transport. The term was introduced by Karl Pearson in 1905 in the context of biological dispersal~\cite{pearson1905problem_1,pearson1905problem_2}, building on Ronald Ross's 1905 studies regarding the spatial spreading of mosquitoes~\cite{ross1905logical,ross1905possibility}; Lord Rayleigh then pointed out that his earlier work already contained the many-step solution~\cite{rayleigh1905problem}. The discrete picture was connected to continuum diffusion by Einstein and Smoluchowski~\cite{einstein1905uber,smoluchowski1906zur,smoluchowski1916drei,gudowska-nowak_preface_2017}. 
Langevin's complementary approach, in which unresolved microscopic degrees of freedom are represented by fluctuating forces, became another central framework for non-equilibrium statistical physics. 
Since then, lattice-based stochastic models have become a versatile framework for diffusion, transport, and collective phenomena, with applications ranging from random walks and percolation~\cite{alexander_density_1982,gefen_anomalous_1983,kesten_subdiffusive_1986,stauffer_introduction_2018,barlow_random_2004,lauber_bonomo_pushy_2026,fisher_random_1984,madras_self-avoiding_1996} to reaction--diffusion processes~\cite{henry2006anomalous,seki2003fractional,perez2019reactive,sleeman1997reaction}, financial markets~\cite{fama_random_1995}, and biological search strategies~\cite{oshanin_efficient_2009,serfozo_optimal_1981,oshanin_intermittent_2007,berg_random_1983,grebenkov2024target}.

The study of systems far from thermal equilibrium has substantially expanded the scope of non-equilibrium statistical physics~\cite{tevrugt2026colloquium,de_gennes_soft_1992}.
One such class of systems is active matter, in which an agent converts energy from an internal source or its local environment into motion through non-conservative forces coupled to its propulsion and orientation~\cite{romanczuk2012active, prawar_dadhichi_origins_2018}. 
Examples of such systems include Brownian motors~\cite{astumian_brownian_2002, reimann_introduction_2002}, motile cells~\cite{fletcher_introduction_2004, koch_collective_2011}, and synthetic self-propelled particles~\cite{bechinger2016active, walther_janus_2008,gomez2016dynamics}.
Unlike passive Brownian particles, active agents typically maintain their direction over a finite persistence time because their propulsion orientation relaxes only through rotational noise or discrete reorientation events such as randomization, reversal, or turning. This directional persistence is a key characteristic of active matter~\cite{obyrne_time_2022}. At larger biological scales, trajectories can additionally exhibit long-range temporal correlations and anomalous transport, as observed in systems extending from molecular motion to migrating birds~\cite{vilk2022unravelling}.
In many systems this directional memory is also chiral, meaning the particle has a preferred handedness, producing circular trajectories, biased turns, and edge-following motion~\cite{lowen2016chirality, liebchen2022chiral}. 
Such chiral active dynamics appear in a broad range of physical and biological systems, including swimming microorganisms~\cite{frymier1995three, lauga2006swimming, petroff2015fast, li2024robust, tan2022odd}, asymmetric self-propelled particles~\cite{kummel2013circular}, chiral fluids~\cite{soni2019odd, massana2021arrested}, and driven active particles~\cite{kummel2013circular, cao2023memory}.
Understanding how persistence and chirality jointly shape transport is therefore a fundamental problem in active matter.

Continuum models such as active Brownian particles (ABPs)~\cite{romanczuk2012active,schimansky1995structure} and active Ornstein--Uhlenbeck particles~\cite{bonilla_active_2019} provide descriptions of persistent active motion. 
In chiral variants of these models, the orientation or active velocity both decorrelates and rotates with a specific handedness, producing exponentially damped oscillatory velocity correlations~\cite{pattanayak2024impact, liebchen2022chiral, caprini2026modeling}.
However, continuum descriptions can obscure the discrete structure of microscopic reorientation events. 
In particular, the distinction between persistence, reversal, left turns, right turns, and random reorientation is often hidden inside coarse-grained rotational diffusion and drift parameters. 
Simple lattice models are therefore valuable because they retain microscopic interpretability, are efficient to simulate at large scales, permit exact calculations, and provide a natural setting for studying transport in structured environments, obstacles, networks, and percolating media.

Persistent motion has a long history in statistical physics, both in the continuum~\cite{taylor1922diffusion,furth1920brownsche} and on lattices~\cite{montroll1950markoff,gilbert2010diffusion,gilbert_diffusion_2012,gilbert_diffusive_2011,gilbert_persistence_2009,bender_correlated_1984,renshaw_correlated_1981,chen_general_1994,larralde1997transport,sadjadi2015persistent}. 
A way to introduce persistence into lattice models is via equipping the walker with an orientational diffusion, which has been studied in one and two dimensions \cite{jose2022active}. A conceptually distinct way to encode persistence is via an internal state that stores the previous direction of motion.
This internal degree of freedom makes the walk Markovian in an enlarged position--orientation space, while producing non-trivial correlations in the projected spatial trajectory. 
By comparison, lattice models with an explicitly chiral internal degree of freedom remain relatively sparse~\cite{osat2026topological}.
Recent work introduced the chiral random walk (CRW) on a lattice~\cite{wojcik2026chiral} using the language of a ``coin'' operator, inspired by discrete-time quantum walks~\cite{aharonov1993quantum}. 
In that formulation, the coin acts on the orientational degree of freedom and determines the stochastic rule by which the walker re-orients itself. 
This provides an intuitive and compact way to encode chirality and memory at the microscopic level.

Here we introduce a minimal active chiral random walk in which persistence, reversal, clockwise and counter-clockwise rotation, and random reorientation are controlled by explicit microscopic probabilities. These microscopic rules reduce, at the level of the first angular harmonic, to the complex parameter \(\lambda=q_p+i q_c\). The real part \(q_p\) controls persistence against reversal, while the imaginary part \(q_c\) controls chirality. This provides a direct connection between microscopic reorientation rules and macroscopic transport.

Relative to the chiral random walk of Ref.~\cite{wojcik2026chiral}, the present construction introduces persistence as a separately tunable property. We use the resulting active chiral random walk (aCRW) to derive the continuum density and polarization equations directly from the stochastic dynamics. We obtain exact directional correlations, the finite-time mean-squared displacement (MSD), and the long-time even and odd transport coefficients. We then establish the asymptotic mapping to the chiral active Brownian particle model and quantify the finite-step corrections away from this limit. The resulting comparison identifies which transport properties are preserved under coarse-graining and which depend on the discrete dynamics. In particular, the leading long-time transport remains exact even when the transient polar dynamics are only approximately represented. Strong chirality and anti-persistent motion define intrinsically discrete regimes beyond the standard smooth angular-diffusion description.
Complementary to Larralde's analysis of chiral persistent transport~\cite{larralde1997transport}, our focus here is the active-matter interpretation and continuum relations made explicit by the intuitive coefficients $q_p$ and $q_c$. 
We use ``active'' here in the kinematic sense of a particle that moves at a fixed self-propulsion speed while its orientation evolves stochastically; the model does not resolve the energetic mechanism generating propulsion.
Further, our model also provides a simple, exactly tractable test of what a continuum approximation preserves. Its correlations and long-time transport follow explicitly from the same microscopic probabilities, allowing us to identify the temporal corrections omitted by the approximation and show that these need not change the leading bulk transport.

This paper is organized as follows. Section~\ref{sec:model} defines the coin-based lattice model and its stochastic interpretation. Section~\ref{sec:continuum_projection} derives the continuum density and polarization equations, including the even and odd currents. Section~\ref{sec:correlations} obtains the exact directional correlations, finite-time MSD, and Green--Kubo transport coefficients. Section~\ref{sec:cABP-connection} compares the aCRW with the continuum chiral active Brownian particle (cABP) model, identifies the weak-turning limit, and distinguishes finite-step correlation matching from transport matching. The appendices collect the directional-correlation derivation (Appendix~\ref{app:correlations}), the polar reduction and hydrodynamic expansion (Appendix~\ref{app:polar-reduction}), the boundary-scattering rules (Appendix~\ref{app:boundary-scattering}), the derivation of the cABP field equation (Appendix~\ref{app:cABP-marginalization}), and the finite-step rate and transport corrections (Appendix~\ref{app:finite-step-matching}).

\section{Model}
\label{sec:model}

Active motile agents often combine stochasticity, persistence, and chirality in their dynamics.
Introducing these properties into a lattice model in an intuitive and controllable manner enables the study of a range of static and dynamical properties, including the evolution and steady states of the displacement probability distribution~\cite{lemaitre2023non, pattanayak2026inertia, zhou2025experimental}, percolation dynamics~\cite{bonomo2023loss, moore2023active, zeitz2017active}, first-passage properties, and search efficiency~\cite{iyaniwura2026mean, baouche2025first,pal2024active-target}, as recently demonstrated for the non-persistent CRW in Ref.~\cite{uhlig2026chirality}. 
We introduce a lattice model that unifies these ingredients within a single coin-based framework while allowing each of these properties to be controlled independently.

Our starting point is the CRW introduced in Ref.~\cite{wojcik2026chiral}, whose formalism we adopt and extend.
There, the walker's displacement is coupled to an internal degree of freedom that stores information about its orientation, and the orientation is propagated by a stochastic operator, the \textit{coin} $\mathrm C$.
The structure of the coin dictates how consecutive steps are correlated and is therefore the object from which the internal dynamical properties of a walker arise in the model. 

\subsection{Microscopic state and update}

The walker occupies a site $(x,y)=\mathbf r\in\ell \mathbb Z^2$ on a two-dimensional square lattice with lattice constant $\ell$, and carries an internal orientation 
\begin{equation}
    d\in D=\{\rightarrow,\, \uparrow,\, \leftarrow,\, \downarrow\},
\end{equation}
which records the direction of its most recent step\footnote{We order $D$ cyclically, which differs by a permutation from the ordering used in Ref.~\cite{wojcik2026chiral}.}. Each internal state is associated with a lattice displacement
\begin{equation}
    \mathbf e_{\rightarrow}=\begin{pmatrix}1\\0\end{pmatrix}, \quad
    \mathbf e_{\uparrow}=\begin{pmatrix}0\\1\end{pmatrix}, \quad
    \mathbf e_{\leftarrow}=\begin{pmatrix}-1\\0\end{pmatrix}, \quad
    \mathbf e_{\downarrow}=\begin{pmatrix}0\\-1\end{pmatrix},
    \label{eq:displacements}
\end{equation}
and we collect these as the columns of the lattice-geometry matrix
\begin{equation}
    \mathrm E  =\begin{pmatrix}
        1 & 0 & -1 & 0\\
        0 & 1 & 0 & -1
    \end{pmatrix},
    \label{eq:geometry-matrix}
\end{equation}
With this notation, $\mathbf e_d = \mathrm E \ket{d}$, where $\ket{\rightarrow}=(1,0,0,0)^\mathrm{T}$, $\ket{\uparrow}=(0,1,0,0)^\mathrm{T}$, $\ket{\leftarrow}=(0,0,1,0)^\mathrm{T}$ and $\ket{\downarrow}=(0,0,0,1)^\mathrm{T}$ are the basis vectors of the internal space.
The joint distribution over configurations is the probability vector 
 \begin{equation}
    \ket{\rho(t)}
    =
    \sum_{\mathbf r, d}
    P_d(\mathbf r,t)\, \ket{\mathbf r}\otimes\ket{d},
    \label{eq:probability-vector}
\end{equation}
where $P_d(\mathbf r,t)=\braket{\mathbf r,d}{\rho(t)}$ is the joint position--orientation probability for the walker to be at $\mathbf r$ with internal state $d$, and $\ket{\mathbf r,d}\equiv\ket{\mathbf r}\otimes\ket{d}$. 
Marginalizing over the internal degrees of freedom returns the spatial probability density function (PDF), $\rho(\mathbf r,t)=\sum_{d\in D}P_d(\mathbf r,t) $. 

In each time step, the walker performs two sequential operations: the coin $\mathrm  C$ updates the internal state, and the conditional-step operator $S$ then displaces the walker along the updated direction,
\begin{equation}
    \ket{\rho(t+\Delta t)}
    =
    \mathrm S( {\mathrm I}_{\mathbf r}\otimes \mathrm  C) \ket{\rho(t)},
    \label{eq:evolution}
\end{equation}
with $\mathrm I_{\mathbf r}$ the identity on the spatial degrees of freedom. 
The step operator is the conditional shift $\mathrm S=\sum_{\mathbf r, d}\ket{\mathbf r+\ell \mathbf e_d, d}\bra{\mathbf r,d},$ so that, for example, $\mathrm S\ket{(x,y),\rightarrow}=\ket{(x+\ell,y),\rightarrow}$.
In other words, the walker reorients according to the coin and moves in its new orientation. The corresponding component master equation is introduced at the start of Sec.~\ref{sec:continuum_projection}; explicit operator matrices and the simulation rule are given below.

\subsection{Coin and angular modes}
The coin is built from five elementary operations: keeping the orientation, reversing it, rotating clockwise (right-handed chirality), rotating counter-clockwise (left-handed chirality), and mixing (randomizing) the orientation. 
Each operator is assigned a fixed probability denoted by the corresponding prefactor, so that 
\begin{equation}
    \mathrm C=q_I \mathrm I+q_R \mathrm R+q_L \mathrm L+q_B \mathrm B+q_M \mathrm M,
    \label{eq:general-coin-main}
\end{equation}
where ${\mathrm I},{\mathrm R},{\mathrm L},{\rm B},{\mathrm  M}$ are the identity, right-chiral, left-chiral, reversal and mixing operators, given explicitly below.
Since the prefactors are probabilities, they are bounded in $[0,1]$ and obey the normalization
\begin{equation}
    \label{eq:normalization}
    q_I+q_R+q_L+q_B+q_M=1 .
\end{equation}
We note that the four non-randomizing operators are mutually orthogonal and obey a symmetric cyclic relation. 
Within the cyclic ordering of $D$, we have $\mathrm R\ket{d_m}=\ket{d_{m-1}}$, where $m=1,2,3,4$ indexes $D$ cyclically (modulo four), and $\mathrm B=\mathrm R^{2}$, $\mathrm L=\mathrm R^{3}=\mathrm R^{-1}$ and $\mathrm R^{4}=\mathrm I$. 
These four matrices therefore form the regular representation of the cyclic group $C_4$, and their real span is the algebra of $4\times4$ circulant matrices. 
The randomization operator is not a fifth independent element, but rather a combination of the four other operators, $\mathrm M=(\mathrm I+\mathrm R+\mathrm B+\mathrm L)/4$, so that Eq.~\eqref{eq:general-coin-main} is an over-parametrization and can be reduced.
Yet we keep this form, as having $\mathrm M$ explicit helps in making the model easier to control and interpret.
For analytical simplification, however, it is useful to present the coin in its minimal form by absorbing $\mathrm M$ as $\mathrm C=\sum_{k=1}^{4}\hat q_k\, \mathrm R^{k}$,
where $ \hat q_i=q_i+q_{M}/4$ for $i\in\{I,R,B,L\}$ are the reduced weights that inherit the normalization from Eq.~\eqref{eq:normalization}. 
We note that by construction, they are exactly the matrix elements of the coin, i.e., $\hat q_I=\bra{d}\mathrm C\ket{d}$, $\hat q_B=\bra{\mathrm B d} \mathrm C\ket{d}$, $\hat q_R=\bra{\mathrm R d}\mathrm  C\ket{d}$, and $\hat q_L=\bra{\mathrm L d}\mathrm  C\ket{d}$. 

Because the coin matrix is circulant, it is diagonalized by the angular Fourier modes of $C_4$, which are 
\begin{equation}
    \ket{v_0}=\begin{pmatrix}1\\1\\1\\1\end{pmatrix}, \quad
    \ket{v_1}=\begin{pmatrix}1\\- \mathrm{i}\\-1\\ \mathrm{i}\end{pmatrix}, \quad
    \ket{v_2}=\begin{pmatrix}1\\-1\\1\\-1\end{pmatrix}, \quad
    \ket{v_3}=\begin{pmatrix}1\\ \mathrm{i}\\-1\\- \mathrm{i}\end{pmatrix}, \quad
    \label{eq:fourier-modes}
\end{equation}
with eigenvalues that follow from $\mathrm C\ket{v_n}=\lambda_n\ket{v_n}$. Explicitly,
\begin{subequations}
\begin{align}
    \lambda_0&=1, &&\text{density mode,}
    \label{eq:lambda0}\\
    \lambda_1&=\left(\hat q_I-\hat q_B\right)+\mathrm{i}\left(\hat q_L-\hat q_R\right)\equiv q_p+\mathrm{i}\,q_c, &&\text{polar mode,}
    \label{eq:lambda1}\\
    \lambda_2&=\left(\hat q_I+\hat q_B\right)-\left(\hat q_R+\hat q_L\right)\equiv q_n, &&\text{nematic mode,}
    \label{eq:lambda2}\\
    \lambda_3&=\left(\hat q_I-\hat q_B\right)-\mathrm{i}\left(\hat q_L-\hat q_R\right) = \overline{\lambda_1}, &&\text{conjugate polar mode.}
    \label{eq:lambda3}
\end{align}
\end{subequations}
Each eigenvalue has a direct microscopic interpretation. The unit density eigenvalue reflects that the coin redistributes probability without creating or destroying it, making the associated PDF $\rho(\mathbf{r},t)$ conserved, as shown in Sec.~\ref{sec:continuum_projection}.
The polar eigenvalue splits into the \textit{persistence}, $q_p=q_I-q_B$, i.e., the excess of keeping the walker orientation ($q_p>0$) over flipping it ($q_p<0$), and the \emph{chirality}, $q_c=q_L-q_R$, i.e., the excess of left ($q_c>0$) over right ($q_c<0$) turns (counter-clockwise rotation is defined positive). 
The nematic eigenvalue $q_n$ measures the preference for collinear ($q_n>0$) over orthogonal ($q_n<0$) steps and is the third free parameter of the coin. 
The three walk parameters $q_p$, $q_c$, and $q_n$ now lie in $[-1,1]$ and inherit shared constraints from the non-negativity of the transition probabilities \eqref{eq:normalization}, \( |q_p|\le (1+q_n)/2\), and \(|q_c|\le (1-q_n)/2\), which projected is \(|q_p|+|q_c|\le 1 \).
Note that $q_p$, $q_c$, and $q_n$ can equally be defined with respect to the original weights $q_i$ instead of the rescaled $\hat q_i$, as $q_M$ cancels in the eigenvalue decomposition.

\subsection{Microscopic operators and stochastic update}
\label{sec:model-details}

The orientation basis and its ordering are defined above. Here we give the explicit operators needed to implement the walk. The step operator acts as
\begin{equation}
\begin{split}
 \mathrm S\ket{(x,y),\rightarrow} &= \ket{(x+\ell,y),\rightarrow},\qquad
 \mathrm S\ket{(x,y),\leftarrow} = \ket{(x-\ell,y),\leftarrow},\\
 \mathrm S\ket{(x,y),\uparrow} &= \ket{(x,y+\ell),\uparrow},\qquad
 \mathrm S\ket{(x,y),\downarrow} = \ket{(x,y-\ell),\downarrow}.
\label{eq:step_operator}
\end{split}
\end{equation}
The five elementary $4\times4$ operators used in the propagation of the direction are
\begin{equation}
\begin{split}
&  \mathrm I=\begin{pmatrix}1&0&0&0\\
0&1&0&0\\
0&0&1&0\\
0&0&0&1\\\end{pmatrix},
~
 \mathrm R=\begin{pmatrix}
0&1&0&0\\
0&0&1&0\\
0&0&0&1\\
1&0&0&0\\
\end{pmatrix},
~
\ \mathrm L=\begin{pmatrix}
0&0&0&1\\
1&0&0&0\\
0&1&0&0\\
0&0&1&0\\
\end{pmatrix},\\
& \mathrm B=\begin{pmatrix}
0&0&1&0\\
0&0&0&1\\
1&0&0&0\\
0&1&0&0\\
\end{pmatrix},
~
\mathrm M=\frac{1}{4}\begin{pmatrix}
1&1&1&1\\
1&1&1&1\\
1&1&1&1\\
1&1&1&1\\
\end{pmatrix}.
\end{split}
\end{equation}
Here $\mathrm I,\mathrm R,\mathrm L,\mathrm B,\mathrm M$ are, respectively, the identity, right-chiral, left-chiral, reversal and mixing (randomization) operators. Their weighted combination is the coin defined in Eq.~\eqref{eq:general-coin-main}.

The coin parameters transfer immediately to the stochastic update rule for the orientation. 
Given a current direction $\ket{d_t}$, the next direction is sampled according to
\begin{equation}
\ket{d_{t+1}}=
\begin{cases}
\ket{d_t}, & \mathrm{prob.}\ q_I,\\
\mathrm B\ket{d_t}, & \mathrm{prob.}\ q_B,\\
\mathrm R\ket{d_t}, & \mathrm{prob.}\ q_R,\\
\mathrm L\ket{d_t}, & \mathrm{prob.}\ q_L,\\
 \ket{d_{\rm unif}},\quad d_{\rm unif}\sim\operatorname{Uniform}(D),
 & \mathrm{prob.}\ q_M.
\end{cases}
\end{equation}
Thus a mixing event samples one of the four orientation states uniformly; \(\mathrm M\ket{d_t}\) is the corresponding probability vector and not itself a realized orientation. 
The associated transition probabilities are the matrix elements of the stochastic coin operator, $(d'|\mathrm C|d)=P(d'|d)$.

\subsection{Representative dynamics}

\begin{figure}[htbp]
    \centering
    \includegraphics[width=1.0\linewidth]{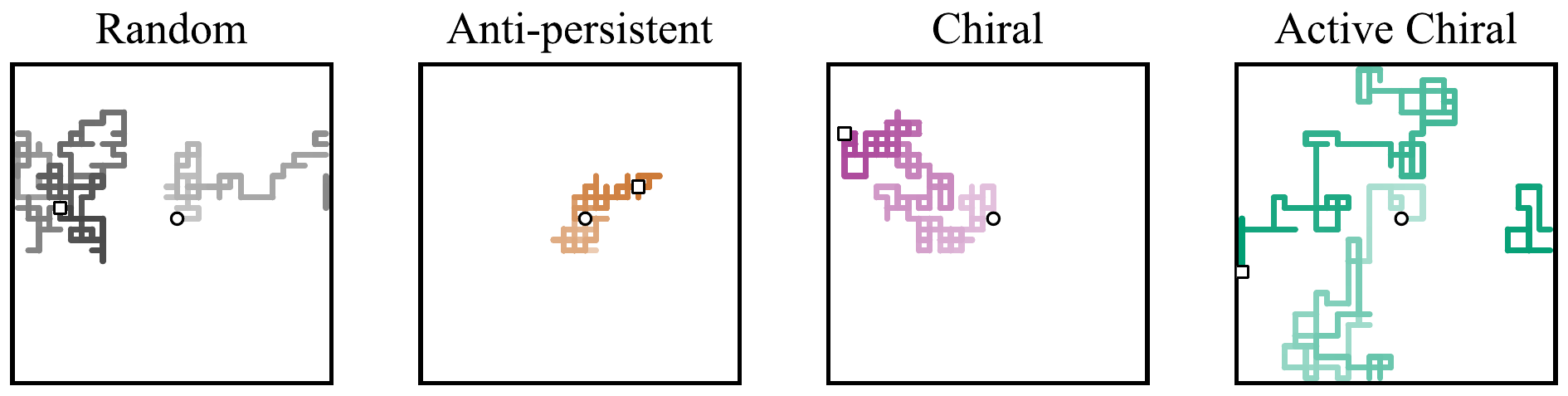}
    \caption{Representative trajectories for four walker types. The starting position is denoted by a circle and the endpoint by a square, while the colour gradient represents the passing of time (from light to dark).
    Each trajectory consists of 300 steps for a single walker.
    Microscopic parameters: random (dark grey), $q_M=1$; anti-persistent (muted orange), $q_B=0.5$, $q_L=0.2$, and $q_M=0.3$; chiral (purple), $q_L=0.5$ and $q_M=0.5$; active chiral (green), $q_I=0.5$, $q_L=0.2$, and $q_M=0.3$.
    The trajectories display the characteristic geometry of each turning rule.}
    \label{fig:Fig1}
\end{figure}

Being a discrete random walk, the aCRW is considerably cheaper to simulate than continuum models of active matter, which is essential when dynamical properties have to be sampled.
The coin in Eq.~\eqref{eq:general-coin-main} also makes the aCRW directly tunable, since every microscopic parameter has an immediate interpretation.
With these two notions in mind, we begin the analysis by observing four different types of walkers: random ($q_M = 1$), anti-persistent ($q_B \neq0$), chiral ($q_R$ or $q_L \neq0$), and active chiral ($q_I \neq0$ and $q_R$ or $q_L \neq0$).
Figure~\ref{fig:Fig1} shows the representative trajectories where different characteristics are directly visible; the compact loops of the chiral walker, the extended runs of the persistent-chiral walker, and the frequent backtracking of the anti-persistent walker. 

This phenomenology raises the question of how these discrete microscopic parameters relate to and control continuum properties. 
We address this independently by both hydrodynamic coarse-graining in Sec.~\ref{sec:continuum_projection} and Green--Kubo calculation in Sec.~\ref{sec:green-kubo}, before comparing the result with established continuum models in Sec.~\ref{sec:cABP-connection}.

\section{From lattice to continuum: density and polarization fields}
\label{sec:continuum_projection}

The detailed description of the model allows us to coarse-grain it to the continuum limit, resulting in a set of equations that describe the non-interacting active chiral gas.
To do so, we begin by projecting Eq.~\eqref{eq:evolution} onto $\bra{\mathbf r,d}$, which yields the orientation-resolved master equation in component form,
\begin{equation}
    P_d(\mathbf r,t+\Delta t)
    =
    \sum_{d'}
    C_{dd'}\, P_{d^\prime} (\mathbf r-\ell \mathbf e_d, t),
    \label{eq:master-equation}
\end{equation}
where $P_d(\mathbf r,t)$ is the joint position--orientation probability for the walker to be at $\mathbf{r}=(x,y)$ with orientation $d\in D$, and $C_{dd'}=\bra d \mathrm C\ket{d'}$ is the conditional probability of updating from orientation $d'$ to $d$.
Note that the displacement is set by the outgoing state $d$ and not by $d'$, since the coin acts before the step.
Summing over all possible orientations gives the spatial PDF,
\begin{equation}
    \rho(\mathbf r,t)=\sum_d P_d(\mathbf r,t).
    \label{eq:marginalized-density}
\end{equation}
The spatial probability is normalized as $\sum_{\mathbf r}\rho(\mathbf r,t)=1$.
Because there is no rest state, all probability at a site is redistributed among the four outgoing orientations at every update. 
Writing the change in density over one time step as $\rho(\mathbf{r},t+\Delta t)-\rho(\mathbf{r},t)$ and replacing this finite difference by a time derivative gives the time evolution for the PDF, where $\nu=1/\Delta t$ is the update frequency,
\begin{equation}
    \frac{\partial \rho(\mathbf r,t)}{\partial t}
    \approx
    \nu
    \left[
    \sum_{d,d'}C_{dd'}\, P_{d^\prime} (\mathbf r-\ell \mathbf e_d, t)
    -
    \rho(\mathbf r,t)
    \right].
    \label{eq:density-master}
\end{equation}
From this dynamical equation, the spatial and orientational moments of the process can be obtained directly.
 
\subsection{Slow fields and polar reduction}
\label{sec:slow_fields}

The resulting lattice-cell PDF in Eq.~\eqref{eq:marginalized-density} is the zeroth orientational moment of \(P_d\). 
In addition, one can retain the leading orientational information, the first orientational moment, which is given by the local polarization
\begin{equation}
\mathbf m(\mathbf r,t)
    =\sum_{d\in D}\mathbf e_d\,P_d(\mathbf r,t)
      =
      \begin{pmatrix}
          P_\rightarrow-P_\leftarrow\\
          P_\uparrow-P_\downarrow
      \end{pmatrix}(\mathbf r,t).
    \label{eq:polarisation}
\end{equation}
Equivalently, in terms of the lattice geometry matrix \(\mathrm E\) introduced in Eq.~\eqref{eq:geometry-matrix}, $\mathbf m = \mathrm E\, (P_\rightarrow,P_\uparrow,P_\leftarrow,P_\downarrow)^{\mathrm T}$.
This marginalization is followed by a hydrodynamic coarse-graining in which \(\rho\) and \(\mathbf m\) are retained as the slow fields. 
The coarse-grained density is not closed under the dynamics since transport between neighbouring cells is controlled by the local polarization which results in an active-matter hierarchy~\cite{cates2013active, te2020relations}. 
At the polar level, however, the four microscopic reorientation processes enter through only two independent combinations; \textit{persistence} and \textit{chirality}. 
As derived in Appendix~\ref{app:polar-reduction}, Eq.~\eqref{eq:coin-intertwining}, the projection of the coin operator onto the real lattice space satisfies
\begin{equation}
     \mathrm E \mathrm C
    =
    \mathrm C_{(2)} \mathrm E,
    \qquad
     \mathrm C_{(2)}
    =
    q_p\,\mathbf 1+q_c\,\boldsymbol\epsilon,
    \label{eq:polar-coin-main}
\end{equation}
where $\mathrm C_{(2)}$ is the real two-dimensional polar representation of the coin and $\boldsymbol\epsilon$ generates a counter-clockwise rotation by $\pi/2$. 
The eigenvalues of $\mathrm C_{(2)}$ are the polar eigenvalue pair $q_p \pm \mathrm i q_c$ which are exactly $\lambda_1$ and $\lambda_3$, defined in Eq.~\eqref{eq:lambda1}.
We note that only these polar eigenvalues enter the projected dynamics. 
Therefore, to simplify, we denote $\lambda\equiv\lambda_1=q_p+\mathrm i q_c
    =|\lambda|\exp(\mathrm i\Omega)$,
so that $|\lambda|=\sqrt{q_p^2+q_c^2}$ encodes the decay of the polar memory and $\Omega=\arg(q_p+iq_c)$ its rotation.

We retain the density and polarization to describe leading bulk diffusion. 
The nematic moment describes alignment along an axis without distinguishing its two opposite directions, and affects the density only beyond the second-order gradient approximation used here~\cite{kalz2024field}. 
Consequently, the transport coefficients depend on $q_p$ and $q_c$, while $q_n$ can still affect short-scale and boundary behaviour. 
The derivation and the lattice-dependent higher-order corrections~\cite{frisch1986lattice, frisch1987lattice} are discussed in Appendix~\ref{app:polar-reduction}.

\subsection{Hydrodynamic equations and closure}

The ordering in gradients is made explicit by expanding Eq.~\eqref{eq:master-equation} in the lattice constant $\ell$. 
Assuming that the shift $\mathbf r-\ell\mathbf e_d$ is small compared with the scale on which $\rho$ and $\mathbf m$ vary, the detailed Taylor expansion and moment projection in Appendix~\ref{app:continuum-derivation} yield the density equation.
To second order in gradients one obtains
\begin{equation}
    \frac{\partial \rho(\mathbf r,t)}{\partial t}
    =
    D_0\nabla^2 \rho(\mathbf r,t)
    -
    \nu\ell\,\nabla\cdot\left[(q_p \mathbf{1}+q_c \boldsymbol{\epsilon})\, \mathbf m(\mathbf r,t)\right]
    +
    \mathcal{O}(\nabla^4),
    \label{eq:density-polarization}
\end{equation}
where $D_0=\ell^2/(4\Delta t)$ is the diffusion coefficient for a memoryless lattice random walk in two dimensions.
To arrive at a closed expression for the density we turn to the dynamics of the polarization field, which obeys, to leading order in gradients,
\begin{equation}
    \frac{\partial\mathbf m(\mathbf r,t)}{\partial t}
    =
    -\nu\left[(1-q_p)\mathbf 1-q_c \boldsymbol{\epsilon}\right]\mathbf m(\mathbf r,t)
    -
    \frac{\nu\ell}{2}\nabla \rho(\mathbf r,t)
    +
    \mathcal{O}(\nabla^2).
    \label{eq:polarization-eq}
\end{equation}
This equation has a direct physical interpretation. In this leading continuous-time description, the polarization relaxes with rate $\nu(1-q_p)$,
while the chiral part rotates the polarization with angular frequency $\nu q_c$.
Thus positive \(q_p\) increases the orientational persistence time, negative \(q_p\) produces anti-persistent reversal, and nonzero \(q_c\) produces chiral rotation of the local polarization.
Their finite-step corrections are discussed in Sec.~\ref{sec:finite-step-generator}.

On diffusive timescales, the density is the slow conserved mode, whereas the polarization is a non-conserved orientational mode with a finite relaxation gap. Neglecting spatial gradients in Eq.~\eqref{eq:polarization-eq} leads to exponential relaxation with characteristic time $\tau_{\mathbf m}=[\nu(1-q_p)]^{-1}$ and rotation frequency $\nu q_c$, whose handedness is dictated by the sign of $q_c$. The exceptional cases $|\lambda|=1$ retain orientational memory indefinitely and thus fall out of the diffusion interpretation.
These limiting cases have a simple dynamical interpretation: $\lambda=1$ yields ballistic motion, $\lambda=-1$ yields deterministic reversal, and $\lambda=\pm\mathrm i$ corresponds to counter-clockwise or clockwise four-step cycles, respectively, the latter having been analyzed in detail in Ref.~\cite{wojcik2026chiral}.

For a density field varying over scales much larger than $\ell$, the density evolves diffusively and therefore more slowly than the polarization it induces. 
The polarization is first order in the density gradient, while its time derivative enters only at the next diffusive order. 
It can consequently be eliminated locally when its relaxation time is short compared with the evolution time of the density field. This standard multiple-scale argument is the basis of the following quasi-stationary closure~\cite{burekovic2026active}.
The orientational relaxation matrix must be invertible, $(1-q_p)^2+q_c^2\neq0$, and the long-wavelength density rate must remain small compared with its relaxation scale. 
The approximation therefore fails near the ballistic limit $q_p\to1$, $q_c\to0$ unless correspondingly longer length and time scales are considered.

These requirements distinguish three steps in the reduction. First, the gradient expansion assumes $\ell|\nabla|\ll1$. Second, the quasi-stationary closure assumes a time-scale separation between the slow density dynamics and the non-conserved polar mode. Third, interpreting the result as the smooth continuous-time limit of an active particle requires the orientational generator $[(1-q_p)\mathbf 1-q_c\boldsymbol\epsilon]/\Delta t$ to remain finite as $\Delta t\to0$, so $1-q_p=\mathcal O(\Delta t)$ and $q_c=\mathcal O(\Delta t)$. We call this the effective \emph{weak-turning} regime of the coarse-grained polar mode. It is microscopically realized by infrequent finite-angle reorientations and is distinct from a genuine small-angular-step limit; Sec.~\ref{sec:weak-turning-limit} returns to this distinction in detail. Outside this scaling, the density equation can still describe the long-time, long-wavelength limit of a discrete walk, but it should not be interpreted as a smooth cABP limit, as we detail in Sec.~\ref{sec:cABP-connection}. The exact discrete Green--Kubo calculation in Sec.~\ref{sec:green-kubo} independently establishes the long-time transport coefficients for $|\lambda|<1$, including regimes in which the continuum-time equation does not resolve the step-scale transient dynamics.

Setting \(\partial\mathbf m/\partial t\approx0\) at leading hydrodynamic order gives the local quasi-stationary polarization
\begin{equation}
    \mathbf m(\mathbf r,t)
    =
    -
    \frac{\ell}{2}
    \frac{(1-q_p)\mathbf 1 +q_c \boldsymbol{\epsilon}}
    {(1-q_p)^2+q_c^2}
    \nabla \rho(\mathbf r,t).
    \label{eq:m-closure}
\end{equation}
Substituting Eq.~\eqref{eq:m-closure} into the density equation, Eq.~\eqref{eq:density-polarization}, produces the closed current--density relation,
\begin{equation}
    \frac{\partial \rho(\mathbf r,t)}{\partial t}=-\nabla\cdot\mathbf j(\mathbf r,t),
    \label{eq:general closed density}
\end{equation}
with
\begin{equation}
    \mathbf j(\mathbf r,t)
    =
    -D_{\rm eff}\nabla \rho(\mathbf r,t)
    -
    D_{\rm odd} \boldsymbol{\epsilon}\nabla \rho(\mathbf r,t).
    \label{eq:current-even-odd}
\end{equation}
The effective and odd diffusion coefficients can then be expressed in terms of the microscopic parameters as
\begin{equation}
    D_{\rm eff}
    =
    D_0
    \frac{1-q_p^2-q_c^2}
    {(1-q_p)^2+q_c^2},
    \qquad
    D_{\rm odd}
    =
    2D_0
    \frac{q_c}
    {(1-q_p)^2+q_c^2},
    \label{eq:effective_diff_coeffs}
\end{equation}
where $D_0=\nu\ell^2/4$ as before.
Equation~\eqref{eq:current-even-odd} separates the regular diffusive current from a chiral current perpendicular to the density gradient. For spatially constant transport coefficients, the latter is divergence-free because $\nabla\cdot(\boldsymbol{\epsilon}\nabla \rho)=0.$
Therefore, the bulk scalar density obeys the effective diffusion equation
\begin{equation}
    \frac{\partial \rho(\mathbf r,t)}{\partial t}
    =
    D_{\rm eff}\nabla^2 \rho(\mathbf r,t).
    \label{eq:closed-density-second-order}
\end{equation}

\begin{figure}[!t]
    \centering
    \includegraphics[width=0.95\linewidth]{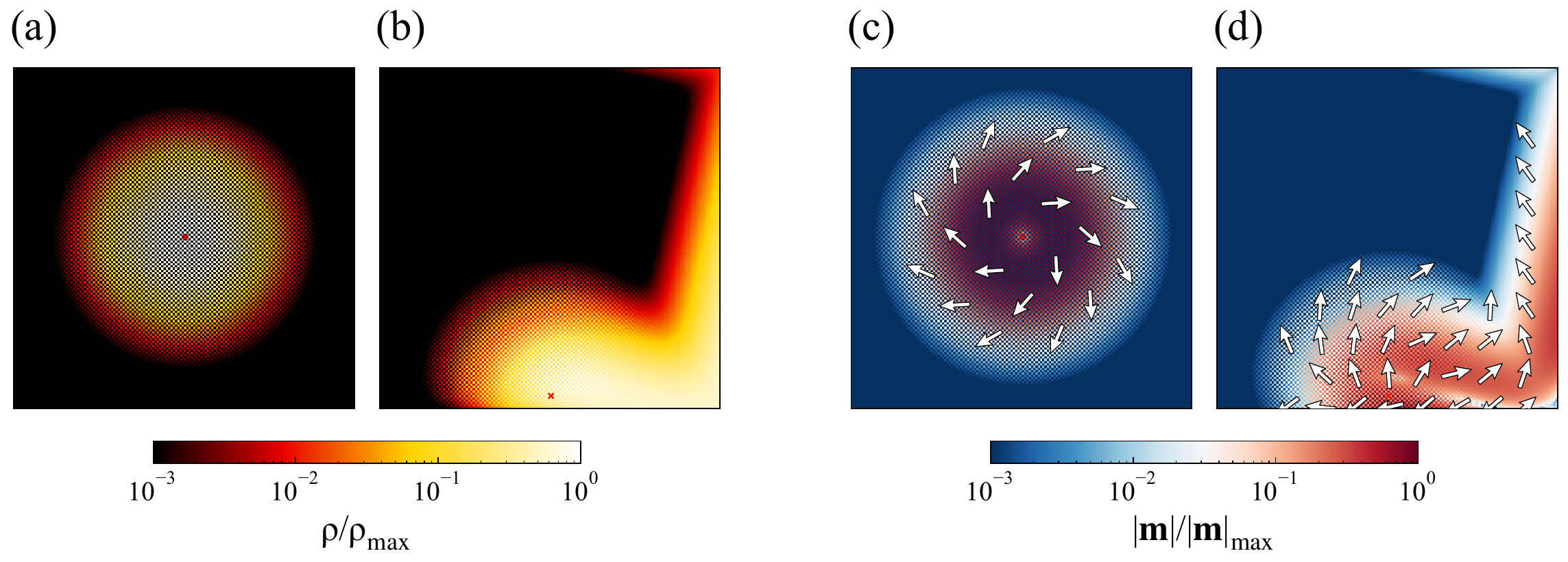}
    \caption{Probability density and orientational polarization of an active chiral random walker with $q_p=0.1$ and $q_c=-0.8$.
    Panels (a,c) show the evolution from a central source, while (b,d) show the evolution from a near-wall source.
    (a,b) Normalized probability density (black--red--yellow--white colour scale).
    (c,d) Direction and normalized magnitude of the local polarization $\mathbf m=\sum_d\mathbf e_d P_d$ (blue--white--red colour scale), showing clockwise orientational circulation and boundary-induced redistribution.
    The near-wall density exhibits a transient wall enhancement, rather than stationary accumulation.
    Both initial conditions occupy a single lattice site with uniform orientations.
    Results are obtained after 2000 updates by exact iteration of Eq.~\eqref{eq:master-equation} on a $150\times150$ lattice, with the specular boundary rule defined in Appendix~\ref{app:boundary-scattering}.}
    \label{fig:density-flux evolution}
\end{figure}

\subsection{Odd current and boundary signature}

The odd diffusion coefficient $D_{\rm odd}$ in Eq.~\eqref{eq:current-even-odd} describes a probability current perpendicular to the density gradient.
This contribution is divergence-free for homogeneous coefficients and therefore does not appear explicitly in the scalar density equation, Eq.~\eqref{eq:closed-density-second-order}, although chirality still affects spreading through $D_{\rm eff}$.
Figure~\ref{fig:density-flux evolution} shows the density and local polarization $\mathbf m$, rather than the probability current.
Because the coin acts on the orientation before displacement, the outgoing vector current is 
\begin{equation}
    \mathbf j_{\rm lat}(\mathbf r,t)
    =
    \nu\ell\left(q_p\mathbf 1+q_c\boldsymbol\epsilon\right)\mathbf m(\mathbf r,t).
    \label{eq:lattice-vector-current}
\end{equation}
The coin can therefore rotate the current relative to the polarization.
The microscopic flux definition and its connection to the continuum current are given in Appendix~\ref{app:microscopic-current}.
The plotted polarization illustrates clockwise orientational circulation, not an independent measurement of transverse current response.

Near a wall, the density and polarization also depend on the microscopic scattering rule and on orientational modes omitted from the bulk closure~\cite{granek2024colloquium,escalante2026benchmarking}.
The redistribution in Fig.~\ref{fig:density-flux evolution} must therefore be interpreted together with the specified boundary rule, rather than as a consequence of $D_{\rm odd}$ alone; the wall dynamics are defined in Appendix~\ref{app:boundary-scattering}.

\section{Velocity correlations}
\label{sec:correlations}

A characteristic signature of chiral dynamics is an antisymmetric off-diagonal correlation between the components of the velocity vector~\cite{kalz2024oscillatory, hargus2021odd}.
We therefore consider the full directional correlation tensor of a freely moving walker.

\subsection{Directional correlations}

As described in Sec.~\ref{sec:model-details}, the conditional-step operator in Eq.~\eqref{eq:step_operator} moves the walker along the lattice direction associated with its updated internal state. Let $d_n$ denote the orientation after the $n$th update. For update frequency $\nu$ and lattice spacing $\ell$, the discrete velocity is $\mathbf v_n=\nu\ell\,\mathbf e_{d_n}$, where $\mathbf e_{d_n}=\mathrm{E}\ket{d_n}$. Apart from the constant factor $\nu\ell$, velocity correlations are therefore directional correlations.

The normalized directional correlation tensor is
\begin{equation}
    \mathbf C_{v}(n)
    =
    2\langle \mathbf e_{d_{m+n}} \otimes \mathbf e_{d_m} \rangle
    =
    C_{v}^{\rm diag}(n)\mathbf 1
    +
    C_{v}^{\rm off}(n)\boldsymbol\epsilon.
    \label{eq:velocity-correlation-tensor}
\end{equation}
Here $C_{v}^{\rm diag}(n)=\langle\mathbf e_{d_{m+n}}\mathbin{\cdot}\mathbf e_{d_m}\rangle$ is the standard diagonal velocity autocorrelation function (VACF), while $C_{v}^{\rm off}(n)=\langle\mathbf e_{d_{m+n}}\mathbin{\cdot}\boldsymbol\epsilon\mathbf e_{d_m}\rangle$ is the off-diagonal correlation.
For an isotropic initial distribution the correlation tensor equals the identity, $\mathbf C_v(0)=\bf 1$.
The conditional-average derivation in Appendix~\ref{app:correlations} shows that successive powers of the projected coin determine both components. 
We find 
\begin{align}
    C_{v}^{\rm diag}(n)=|\lambda|^n\cos(n\Omega)=\operatorname{Re}\lambda^n,
    \label{eq:VACF}
    \\
\intertext{and}
    C_{v}^{\rm off}(n)=|\lambda|^n\sin(n\Omega)=\operatorname{Im}\lambda^n,
    \label{eq:VCCF}
\end{align}
where $\lambda = |\lambda|\mathrm{exp}(\mathrm{i}\Omega)$ is the persistence-chirality parameter defined in Sec.~\ref{sec:slow_fields}.
Essentially, the coin correlates consecutive steps through a constant complex one-step factor, producing an exponentially decaying memory for $|\lambda|<1$ and oscillations when $\Omega\neq0$. 
These characteristics appear in the diagonal and off-diagonal correlations in Fig.~\ref{fig:fitting MSD and VCF}(c,d).
This makes explicit the connection between chirality and oscillatory directional correlations. 
In the model they are directly governed by intuitive microscopic constants $q_p$ and $q_c$.

We note that exponentially damped and oscillatory velocity correlations are also characteristic of active-chiral systems such as cABPs~\cite{pattanayak2024impact}.

\subsection{Exact finite-time mean-squared displacement}

Now that we have identified the velocity correlations, we can calculate the exact
finite-time MSD. After \(N\) steps, the displacement is
\begin{equation}
    \mathbf r_N-\mathbf r_0
    =
    \ell\sum_{j=1}^{N}\mathbf e_{d_j}.
\end{equation}
We denote the mean-squared displacement by $\operatorname{MSD}_N=\langle|\mathbf r_N-\mathbf r_0|^2\rangle$.
Using stationarity of the step increments and
\(C_v^{\rm diag}(n)=\operatorname{Re}(\lambda^n)\), we obtain
\begin{equation}
    \left\langle
    \left|\mathbf r_N-\mathbf r_0\right|^2
    \right\rangle
    =
    \ell^2
    \left[
    N+
    2\sum_{n=1}^{N-1}(N-n)C_{v}^{\rm diag}(n)
    \right]=
    \ell^2
    \left[
    N+
    2\operatorname{Re}
    \sum_{n=1}^{N-1}(N-n)\lambda^n
    \right].
    \label{eq:MSD-exact-sum}
\end{equation}
For \(\lambda\neq1\), the geometric sum can be evaluated explicitly, leading to
\begin{equation}
    \left\langle
    \left|\mathbf r_N-\mathbf r_0\right|^2
    \right\rangle
    =
    \ell^2
    \left[
    N+
    2\operatorname{Re}
    \left(
    \frac{
    \lambda\left[(N-1)-N\lambda+\lambda^N\right]
    }{
    (1-\lambda)^2
    }
    \right)
    \right].
    \label{eq:MSD-exact}
\end{equation}
Equation~\eqref{eq:MSD-exact} describes both the transient and long-time
behaviour of the walk. It remains valid for $\lambda=-1$ and $\lambda=\pm\mathrm i$; only $\lambda=1$ must be evaluated directly from Eq.~\eqref{eq:MSD-exact-sum}. The corresponding special cases are
\begin{equation}
\operatorname{MSD}_N=
\begin{cases}
\ell^2N^2, & \lambda=1,\\
0\ \text{for even }N,\quad \ell^2\ \text{for odd }N, & \lambda=-1,\\
0,\ \ell^2,\ 2\ell^2,\ \ell^2
\quad\text{for }(N\bmod 4)=0,1,2,3, & \lambda=\pm\mathrm i.
\end{cases}
\label{eq:MSD-special-cases}
\end{equation}
Thus, the ballistic case grows quadratically, while deterministic reversal and deterministic chiral rotation remain spatially bounded. For \(|\lambda|<1\), the leading large-\(N\) contribution
is
\begin{equation}
    \left\langle
    \left|\mathbf r_N-\mathbf r_0\right|^2
    \right\rangle
    \sim
    \ell^2N
    \left[
    1+
    2\operatorname{Re}
    \left(
    \frac{\lambda}{1-\lambda}
    \right)
    \right]
    =
    4D_{\rm eff}t,
    \label{eq:MSD-long-time}
\end{equation}
where $t=N\Delta t$.
The long-time diffusion form, unlike the exact finite-time expression, requires $|\lambda|<1$. 
At $|\lambda|=1$, $\lambda=1$ gives ballistic motion, $\lambda=-1$ deterministic reversal, and $\lambda=\pm\mathrm i$ counter-clockwise or clockwise four-step cycles respectively.

\subsection{Green--Kubo transport coefficients}
\label{sec:green-kubo}

The full even and odd transport tensor follows directly from a one-sided discrete Green--Kubo relation~\cite{wojcik2026chiral, gilbert2010diffusion}, without the quasi-stationary closure or an assumption of slowly varying fields,
\begin{equation}
    {\mathbf D}^{\rm GK}
    =
    \frac{\nu\ell^2}{4}
    \left[
        \mathbf C_v(0)
        +
        2\sum_{n=1}^{\infty}\mathbf C_{v}(n)
    \right]
    =D_{\rm eff}^{\rm GK}\mathbf 1+D_{\rm odd}^{\rm GK}\boldsymbol\epsilon.
    \label{eq:GK-discrete}
\end{equation}
The effective diffusion coefficient determines the long-time scalar spreading, whereas the odd-diffusion coefficient cancels from the MSD and the homogeneous scalar diffusion equation but remains in the current--density relation, Eq.~\eqref{eq:current-even-odd}. 
For $|\lambda|<1$, evaluating the geometric sum $\sum_{n=1}^{\infty}\lambda^n=\lambda/(1-\lambda)$, together with $
\mathbf C_v(0)=\bf 1$, leads to
\begin{align}
    D_{\rm eff}^{\rm GK}
    &=
    \frac{\nu\ell^2}{4}
    \left[
    1+2\operatorname{Re}\left(\frac{\lambda}{1-\lambda}\right)
    \right]
    =
    \frac{\nu\ell^2}{4}
    \frac{1-q_p^2-q_c^2}{(1-q_p)^2+q_c^2},
    \label{eq:Deff-GK}\\
    D_{\rm odd}^{\rm GK}
    &=
    \frac{\nu\ell^2}{2}
    \operatorname{Im}\left(\frac{\lambda}{1-\lambda}\right)
    =
    \frac{\nu\ell^2}{2}
    \frac{q_c}{(1-q_p)^2+q_c^2}.
    \label{eq:Dodd-GK}
\end{align}
The diagonal and off-diagonal correlations therefore determine the effective and odd current coefficients, respectively. In particular, $D_{\rm odd}^{\rm GK}$ changes sign under $q_c\to-q_c$ and vanishes in the absence of chirality, $q_c=0$. Equations~\eqref{eq:Deff-GK} and \eqref{eq:Dodd-GK} are identical to the hydrodynamic coefficients in Eq.~\eqref{eq:effective_diff_coeffs}: $D_{\rm eff}^{\rm GK}=D_{\rm eff}$ and $D_{\rm odd}^{\rm GK}=D_{\rm odd}$. Thus the two calculations retain the same leading long-time transport information, even though the hydrodynamic equation applies only after its additional scale-separation assumptions.
This preservation of lattice transport, however, does not imply exact agreement with cABP correlations or transport at finite $\Delta t$, as discussed in Sec.~\ref{sec:cABP-connection}.

\begin{figure}[!htb]
    \centering
    \includegraphics[width=1.0\linewidth]{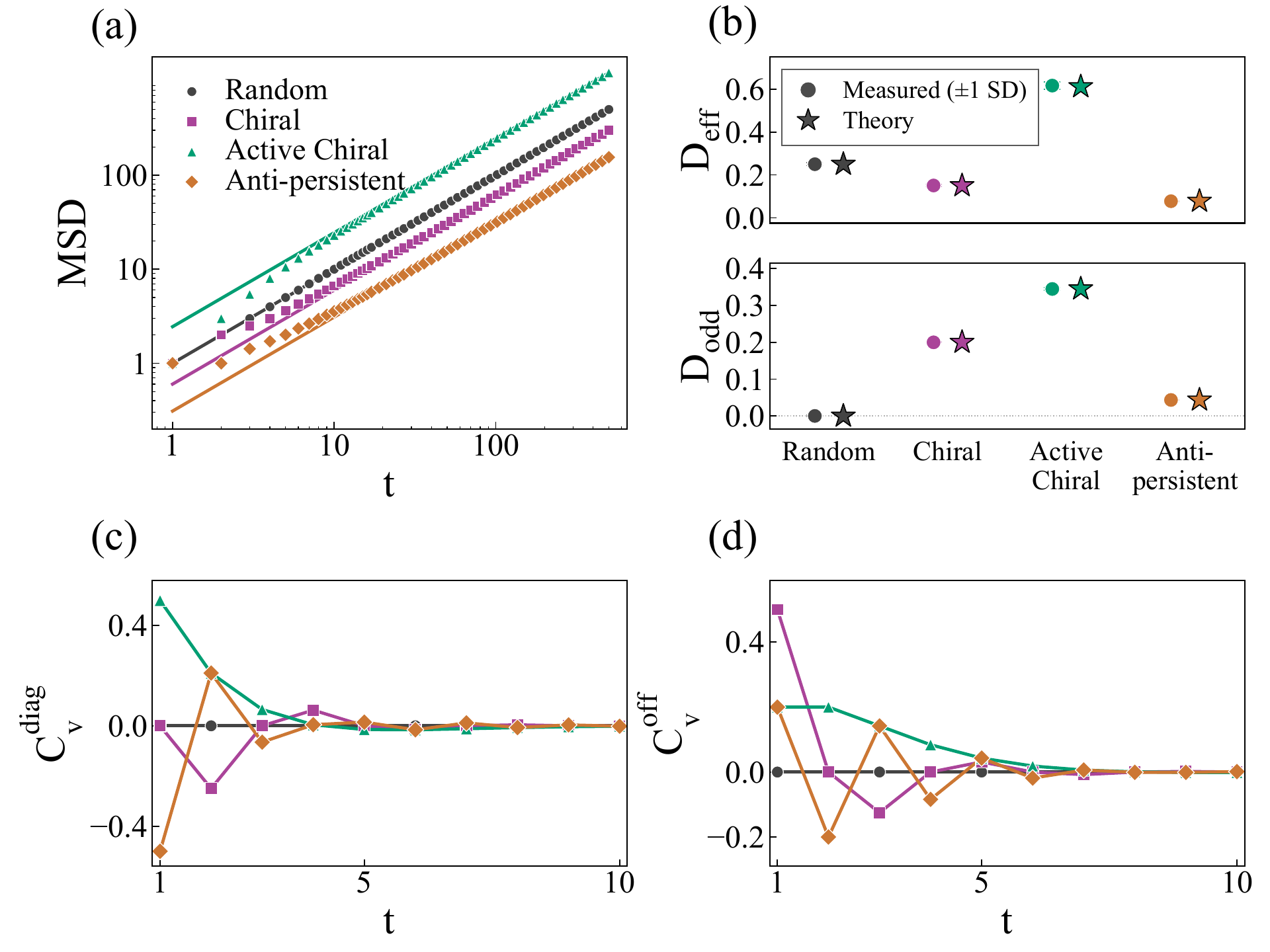}
\caption{Simulation tests of the predicted correlations and transport coefficients.
(a) Measured MSDs (markers) and long-time predictions $\mathrm{MSD}=4D_{\rm eff}t$ (solid lines), using Eq.~\eqref{eq:effective_diff_coeffs}.
(b) Diffusion-coefficient estimates (circles) and theoretical values (stars). $D_{\rm eff}$ is extracted from the MSD slope over $100\leq t\leq500$, and $D_{\rm odd}$ from measured off-diagonal correlations using the discrete Green--Kubo as described in Section~\ref{sec:green-kubo}. 
Markers show the mean across batches; error bars indicate $\pm1$ standard deviation across 50 independent batches of 2000 trajectories.
(c,d) Diagonal and off-diagonal velocity correlations: simulations (markers) and predictions from Eqs.~\eqref{eq:VACF} and \eqref{eq:VCCF} (solid lines).
In panels (a,c,d), dark-grey circles represent the random walker with $(q_p,q_c)=(0,0)$, purple squares the chiral walker with $(0,0.5)$, green triangles the active chiral walker with $(0.5,0.2)$, and muted-orange diamonds the anti-persistent walker with $(-0.5,0.2)$.
Each ensemble contains $10^5$ trajectories evolved for 500 timesteps, with $\ell=\Delta t=1$.}
    \label{fig:fitting MSD and VCF}
\end{figure}

Figure~\ref{fig:fitting MSD and VCF} tests these predictions: panel (a) compares the measured MSD with Eq.~\eqref{eq:MSD-long-time}; panel (b) compares coefficient estimates with Eq.~\eqref{eq:effective_diff_coeffs}; and panels (c,d) compare the diagonal and off-diagonal correlations with Eqs.~\eqref{eq:VACF} and \eqref{eq:VCCF}.
Note that the odd-diffusion-coefficient comparison is a trajectory-based Green--Kubo check, not a separate measurement of transverse response to an imposed density gradient.
The simulations agree closely with the predicted long-time MSD slopes and exact correlation functions.
For each batch, $C_v^{\rm off}(n)$ is estimated by averaging $\mathbf e_{d_{m+n}}\mathbin{\cdot}\boldsymbol\epsilon\mathbf e_{d_m}$ over trajectories and all available time origins. The $D_{\rm odd}$ estimate is $\nu\ell^2/2$ times the sum of these measured correlations over lags $1$--$20$, which, as one can note in panels (c,d), is sufficient.
The numerical implementation of the model and figure generation is openly available in Ref.~\cite{sherf2026aCRWcodes-v120}.

\subsection{Enhanced, reduced, and intrinsically discrete regimes}

\begin{figure}[htbp]
    \centering
    \includegraphics[width=0.65\linewidth]{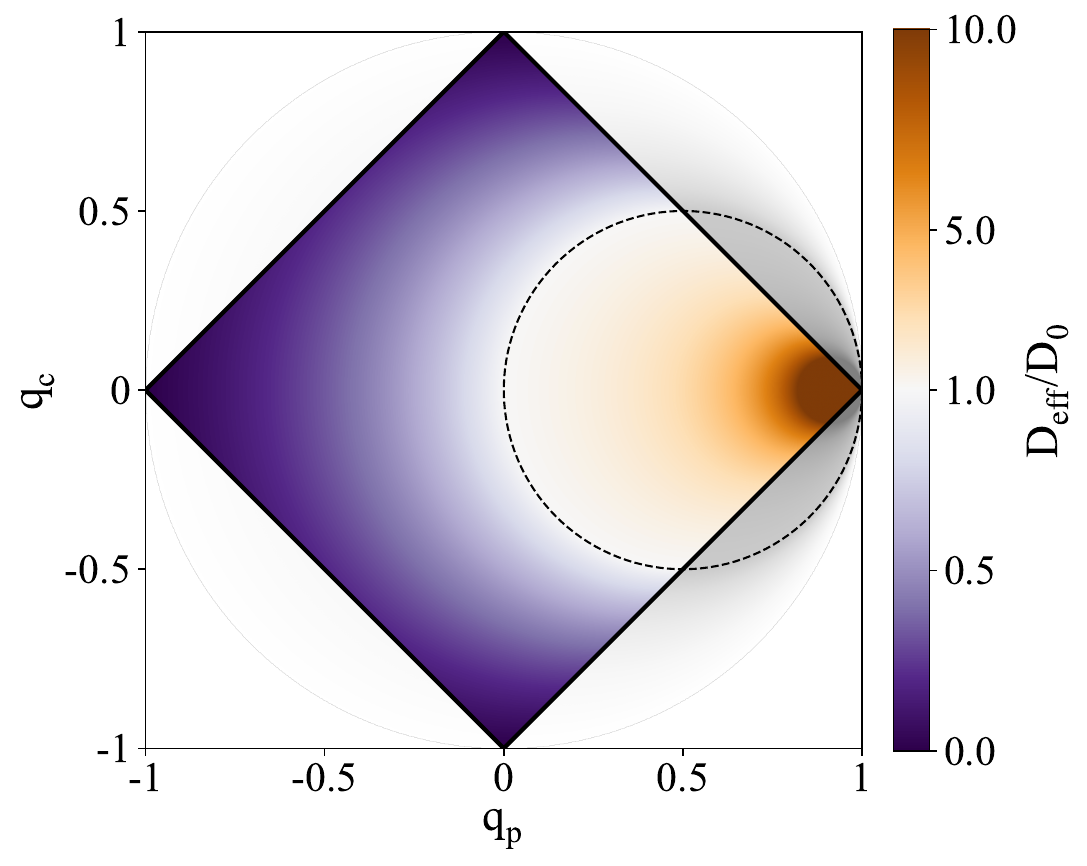}
    \caption{Ratio of the effective diffusion coefficient in Eq.~\eqref{eq:Deff-GK} to the free diffusion coefficient $D_0=\ell^2/(4\Delta t)$ as a function of persistence $q_p$ and chirality $q_c$. The dashed black circle is the boundary $D_{\rm eff}/D_0=1$ in Eq.~\eqref{eq:enhancement-boundary}; it separates enhanced from reduced long-time diffusion, which is also specifically accounted for in the two-scale colour-coding of the figure. The solid black diamond is the boundary imposed by the normalization condition, $|q_p|+|q_c|\le1$; the region outside it is inaccessible.}
    \label{fig:enhanced/reduced diffusion}
\end{figure}

Normalizing Eq.~\eqref{eq:Deff-GK} by the free lattice diffusion coefficient $D_0=\ell^2/(4\Delta t)$ identifies enhanced ($D_{\rm eff}>D_0$) and reduced ($D_{\rm eff}<D_0$) long-time diffusion. 
The boundary between enhanced and reduced diffusion is the circle
\begin{equation}
    \left(q_p-\frac12\right)^2+q_c^2=\frac14,
    \label{eq:enhancement-boundary}
\end{equation}
restricted by normalization to the physically accessible diamond $|q_p|+|q_c|\le1$, as shown in Fig.~\ref{fig:enhanced/reduced diffusion}. 
Hence the model provides a direct microscopic route for tuning the walker between persistence-enhanced and chirality- or reversal-suppressed diffusion.
Further analysis of Eq.~\eqref{eq:Deff-GK} separates the system into distinct dynamical regimes. For \(q_c=0\),
\begin{equation}
    \frac{D_{\rm eff}}{D_0}
    =
    \frac{1+q_p}{1-q_p}.
\end{equation}
Thus \(q_p>0\) enhances diffusion by persistence, while \(q_p<0\) suppresses diffusion by anti-persistence. 
In the limit \(q_p\to-1\), the walker repeatedly reverses direction and the long-time diffusion vanishes. 
This regime is intrinsically discrete and does not correspond to the usual smooth ABP limit. 
Indeed, for \(q_p<0\) and \(q_c=0\), one has \(\lambda=-|\lambda|\), or equivalently \(\Omega= \pi\), far outside the weak-turning continuum condition imposed by the quasi-stationary closure~\cite{wojcik2026chiral}.

For \(q_p=0\), chirality alone gives
\begin{equation}
    \frac{D_{\rm eff}}{D_0}
    =
    \frac{1-q_c^2}{1+q_c^2}.
\end{equation}
Thus, pure chirality reduces the scalar diffusion coefficient by bending trajectories into rotating motion. 
At the deterministic chiral point \(|q_c|=1\), the walker circulates around elementary lattice four-step loops and the long-time scalar diffusion vanishes, even though the dynamics remains strongly chiral~\cite{wojcik2026chiral}.
We note that the scalar diffusion coefficient measures spatial spreading and not the full degree of activity or irreversibility. In particular, the handedness of the walker does not affect it, as $q_c$ enters only quadratically in Eq.~\eqref{eq:Deff-GK}, contrary to the odd-diffusion coefficient in Eq.~\eqref{eq:Dodd-GK} where the sign matters.

\section{Connection to the chiral active Brownian particle model}
\label{sec:cABP-connection}

Motivated by the connections identified throughout the manuscript, this section compares, maps, and distinguishes the aCRW and cABP models.
Specifically, the polar mode of the aCRW displays the relaxation and rotation characteristic of persistent chiral particles, as prominently studied in the cABP model~\cite{liebchen2022chiral,pattanayak2024impact,kalz2024oscillatory}. 
Here we compare the long-time even and odd transport coefficients and correlations with those of a cABP. 
The comparison is made first in the weak-turning limit, where the models best match. 
We then identify the temporal truncation behind this leading mapping and explain why exact matching of sampled velocity correlations does not imply exact transport matching.

\subsection{Continuum cABP model}

We consider a standard two-dimensional cABP with position \(\mathbf r(t)\) and orientation \(\mathbf u(\phi)=(\cos\phi,\sin\phi)\)~\cite{risken1989fokker,pattanayak2024impact}. Its overdamped Langevin dynamics are
\begin{equation}
d\mathbf r
=
\tilde v_0\mathbf u\,dt
+
\sqrt{2\tilde D}\,d\mathbf B,
\qquad
d\phi
=
\tilde\omega\,dt
+
\sqrt{2\tilde D_r}\,dW.
\label{eq:cabp-langevin}
\end{equation}
Here \(\tilde v_0\) is the self-propulsion speed, \(\tilde D\) is an independent translational diffusion coefficient, \(\tilde D_r\) is the rotational diffusion coefficient, and \(\tilde\omega\) is the intrinsic angular velocity. We denote $\mathbf B$ and $W$ are standard Wiener processes. 
We use tildes to distinguish the cABP coefficients from those of the discrete model. 
Throughout, positive chirality denotes counter-clockwise rotation: $\tilde\omega>0$ produces counter-clockwise angular drift, and $\boldsymbol\epsilon$ represents a counter-clockwise rotation by $\pi/2$.

The corresponding Fokker--Planck equation for the angular PDF \(\tilde p(\mathbf r,\phi,t)\) is
\begin{equation}
\frac{\partial \tilde p(\mathbf r,\phi,t)}{\partial t}
=
\tilde D\nabla^2 \tilde p
-
\tilde v_0\mathbf u\cdot\nabla \tilde p
+
\tilde D_r\frac{\partial^2 \tilde p}{\partial\phi^2}
-
\tilde\omega\frac{\partial \tilde p}{\partial\phi}.
\label{eq:cabp-fokker-planck}
\end{equation}
The cABP PDF and polarization field are

\begin{align}
\tilde\rho(\mathbf r,t)
&=
\int_0^{2\pi} \tilde p(\mathbf r,\phi,t)\,d\phi,
\\
\intertext{and}
\tilde{\mathbf m}(\mathbf r,t)
&=
\int_0^{2\pi}\mathbf u(\phi) \tilde p(\mathbf r,\phi,t)\,d\phi.
\end{align}
Following standard projection schemes~\cite{cates2013active, te2020relations}, we apply a first-harmonic, long-wavelength closure in which the nematic moment is neglected. This yields
\begin{align}
\frac{\partial\tilde\rho(\mathbf r,t)}{\partial t}
&=
\tilde D\nabla^2\tilde\rho(\mathbf r,t)
-
\tilde v_0\nabla\cdot\tilde{\mathbf m}(\mathbf r,t),
\label{eq:cabp-density}
\\
\intertext{and}
\frac{\partial\tilde{\mathbf m}(\mathbf r,t)}{\partial t}
&=
\tilde D\nabla^2\tilde{\mathbf m}(\mathbf r,t)
-
\tilde D_r\tilde{\mathbf m}(\mathbf r,t)
+
\tilde\omega\boldsymbol{\epsilon}\tilde{\mathbf m}(\mathbf r,t)
-
\frac{\tilde v_0}{2}\nabla\tilde\rho(\mathbf r,t).
\label{eq:cabp-polarization}
\end{align}
The factor $1/2$ in the last term follows from the isotropic part of the second angular moment. The complete marginalization is shown in Appendix~\ref{app:cABP-marginalization}. In the slow-density, fast-polarization regime, quasi-stationary elimination of $\tilde{\mathbf m}$ gives
\(
\tilde{\mathbf j}=-\tilde D_{\mathrm{eff}}\nabla\tilde\rho-\tilde D_{\mathrm{odd}}\boldsymbol{\epsilon}\nabla\tilde\rho,
\)
where
\begin{equation}
\tilde D_{\mathrm{eff}}
=
\tilde D
+
\frac{\tilde v_0^2\tilde D_r}
{2\left(\tilde D_r^2+\tilde\omega^2\right)},
\qquad
\tilde D_{\mathrm{odd}}
=
\frac{\tilde v_0^2\tilde\omega}
{2\left(\tilde D_r^2+\tilde\omega^2\right)}.
\label{eq:cabp-diffusion-coefficients}
\end{equation}
Thus, the spatial diffusion coefficient $\tilde D$ contributes only to the even coefficient, while $\tilde D_{\mathrm{odd}}$ is generated entirely by the coupling between propulsion and persistent angular rotation. Positive $\tilde\omega$ gives $\tilde D_{\mathrm{odd}}>0$, consistently with the counter-clockwise convention.

\subsection{Weak-turning transport correspondence}
\label{sec:weak-turning-limit}

We now compare these coefficients with the exact aCRW coefficients in Eq.~\eqref{eq:effective_diff_coeffs}. The relevant continuum limit keeps the cABP rates and speed fixed while the discrete time and length scales vanish together
\begin{equation}
q_p
=
1-\tilde D_r\Delta t
+
\mathcal O(\Delta t^2),
\qquad
q_c
=
\tilde\omega\Delta t
+
\mathcal O(\Delta t^2),
\qquad
\ell
=
\tilde v_0\Delta t.
\label{eq:weak-turning-scaling}
\end{equation}
At finite $\Delta t$, the corresponding leading-order identification used below is
\begin{equation}
\tilde D_r=\nu(1-q_p),
\qquad
\tilde\omega=\nu q_c,
\qquad
\tilde v_0=\nu\ell.
\label{eq:relating coefficients}
\end{equation}
In this scaling, the lattice reference coefficient vanishes,
\begin{equation}
D_0
=
\frac{\ell^2}{4\Delta t}
=
\frac{\tilde v_0^2\Delta t}{4}
\xrightarrow{\Delta t\to0}0.
\label{eq:D0-weak-limit}
\end{equation}
The vanishing of the aCRW coefficient \(D_0\) in the continuum limit is expected, since it arises from the finite lattice step size. To compare the two models on equal footing, we consider a cABP without independent translational Brownian diffusion, \(\tilde D=0\), and compare only the contributions arising from the active dynamics.

Substitution of Eq.~\eqref{eq:weak-turning-scaling} into the aCRW coefficients gives
\begin{equation}
\begin{aligned}
D_{\mathrm{eff}}
&\xrightarrow{\Delta t\to0}
\frac{\tilde v_0^2\tilde D_r}
{2\left(\tilde D_r^2+\tilde\omega^2\right)}
=\left.\tilde D_{\mathrm{eff}}\right|_{\tilde D=0},\\
D_{\mathrm{odd}}
&\xrightarrow{\Delta t\to0}
\frac{\tilde v_0^2\tilde\omega}
{2\left(\tilde D_r^2+\tilde\omega^2\right)}
=\tilde D_{\mathrm{odd}}.
\end{aligned}
\label{eq:weak-turning-transport-limit}
\end{equation}
The even and odd long-time transport coefficients of the aCRW therefore approach those of a cABP without independent translational diffusion. This establishes an asymptotic transport correspondence between the models in the weak-turning limit.

The stochastic coin restricts which cABP rates can be reached by this four-direction construction. Near $q_p=1$, the physical domain $|q_p|+|q_c|\leq1$ becomes $|q_c|\leq1-q_p$. With Eq.~\eqref{eq:weak-turning-scaling}, this yields \(|\tilde\omega|\leq\tilde D_r.\)
In this limit, the aCRW reorientation events become infrequent, although the walker still changes direction through a finite angle of $90^\circ$. This differs from the cABP, for which the change in orientation becomes infinitesimally small. 
We therefore use \emph{weak turning} for the effective coarse-grained polar dynamics and \emph{infrequent reorientation} for its microscopic realization. 
A genuine small-angular-step limit would instead require an orientational lattice with \(z\) directions and \(z\to\infty\)~\cite{larralde1997transport}.

\subsection{Finite-step correlations and the continuum approximation}
\label{sec:finite-step-generator}

This comparison connects the intuitive active-matter coefficients $q_p$ and $q_c$ to cABP relaxation and rotation rates, while making the finite-step approximation explicit.
The underlying relation is general: for any mode multiplied by a nonzero factor $\lambda$ at each update, its exponential continuous-time rate is $\log\lambda/\Delta t$.
In the aCRW, the factor is simply $\lambda=q_p+\mathrm iq_c$, so the exact rate and its continuum approximation follow directly from the microscopic probabilities.
The aCRW correlations contain the exact factor $\lambda^n$, whereas the cABP first angular harmonic evolves as $\exp[(-\tilde D_r+\mathrm i\tilde\omega)t]$.
For $0<|\lambda|<1$, matching them at $t=n\Delta t$ gives
\begin{equation}
    \tilde D_r^{\rm corr}
    =-\nu\ln|\lambda|
    =-\nu\ln\sqrt{q_p^2+q_c^2},
    \qquad
    \tilde\omega^{\rm corr}=\nu\arg\lambda.
    \label{eq:exact-correlation-mapping}
\end{equation}
The phase is chosen continuously near $\lambda=1$.

The leading continuum map in Eq.~\eqref{eq:relating coefficients} instead uses $(\lambda-1)/\Delta t=-\tilde D_r+\mathrm i\tilde\omega$.
Their relationship follows directly from
\begin{equation}
    \log\lambda
    =(\lambda-1)-\frac{(\lambda-1)^2}{2}
    +\mathcal O\!\left(|\lambda-1|^3\right).
    \label{eq:modal-generator-truncation}
\end{equation}
The leading map therefore retains only the first term of the exact correlation-rate expansion.
The omitted corrections vanish in the weak-turning limit and are calculated in Appendix~\ref{app:finite-step-generator}.

The simplicity of the model makes the omitted temporal corrections and preserved long-time transport explicit.
The homogeneous polar projection in Eq.~\eqref{eq:polar-coin-main} preserves $\lambda$; replacing a finite time difference by a derivative approximates its evolution, rather than discarding the polar mode.
This differs from neglecting higher angular moments.
Nevertheless, the hydrodynamic coefficients coincide with the exact long-time lattice coefficients obtained in Sec.~\ref{sec:green-kubo}.
The aCRW therefore gives an explicit microscopic example of a general distinction: preserving long-time transport does not require reproducing the exact transient dynamics of the retained mode.

\subsection{Consequences for transport}
\label{sec:correlation-transport-matching}

Matching correlations at update times does not guarantee matching transport coefficients: the lattice Green--Kubo relation sums correlations, whereas the cABP relation integrates them over continuous time.
Even identical sampled values need not give the same sum and integral.
Appendix~\ref{app:finite-step-transport} shows that the difference under correlation matching is of order $\Delta t^2$ at fixed correlation-matched rates and speed.

For comparison, evaluating the cABP coefficients with the leading map, equal speeds $\tilde v_0=\ell/\Delta t$, and $\tilde D=0$ gives
\begin{equation}
    \tilde D_{\rm eff}-D_{\rm eff}=D_0,
    \qquad
    \tilde D_{\rm odd}=D_{\rm odd}.
    \label{eq:leading-map-transport-offset}
\end{equation}
Thus the leading map has an exact offset in the even coefficient, but reproduces the odd coefficient.
At fixed speed, $D_0=\tilde v_0^2\Delta t/4$ vanishes in the continuum limit.

The microscopic distinction is also visible in the current.
The cABP active current is $\tilde v_0\tilde{\mathbf m}$, whereas the aCRW current contains $(q_p\mathbf 1+q_c\boldsymbol\epsilon)\mathbf m$: the coin acts directly on the outgoing flux during each update.
This finite-step correction disappears in the weak-turning limit.
Strong chirality and anti-persistence remain natural aCRW regimes outside the standard cABP correspondence, although related dynamics can be represented by velocity-jump, run-and-tumble, and telegraph models~\cite{othmer1988models,tailleur2008statistical, goldstein1951on}.

\begin{figure}[htbp]
    \centering
    \includegraphics[width=1\linewidth]{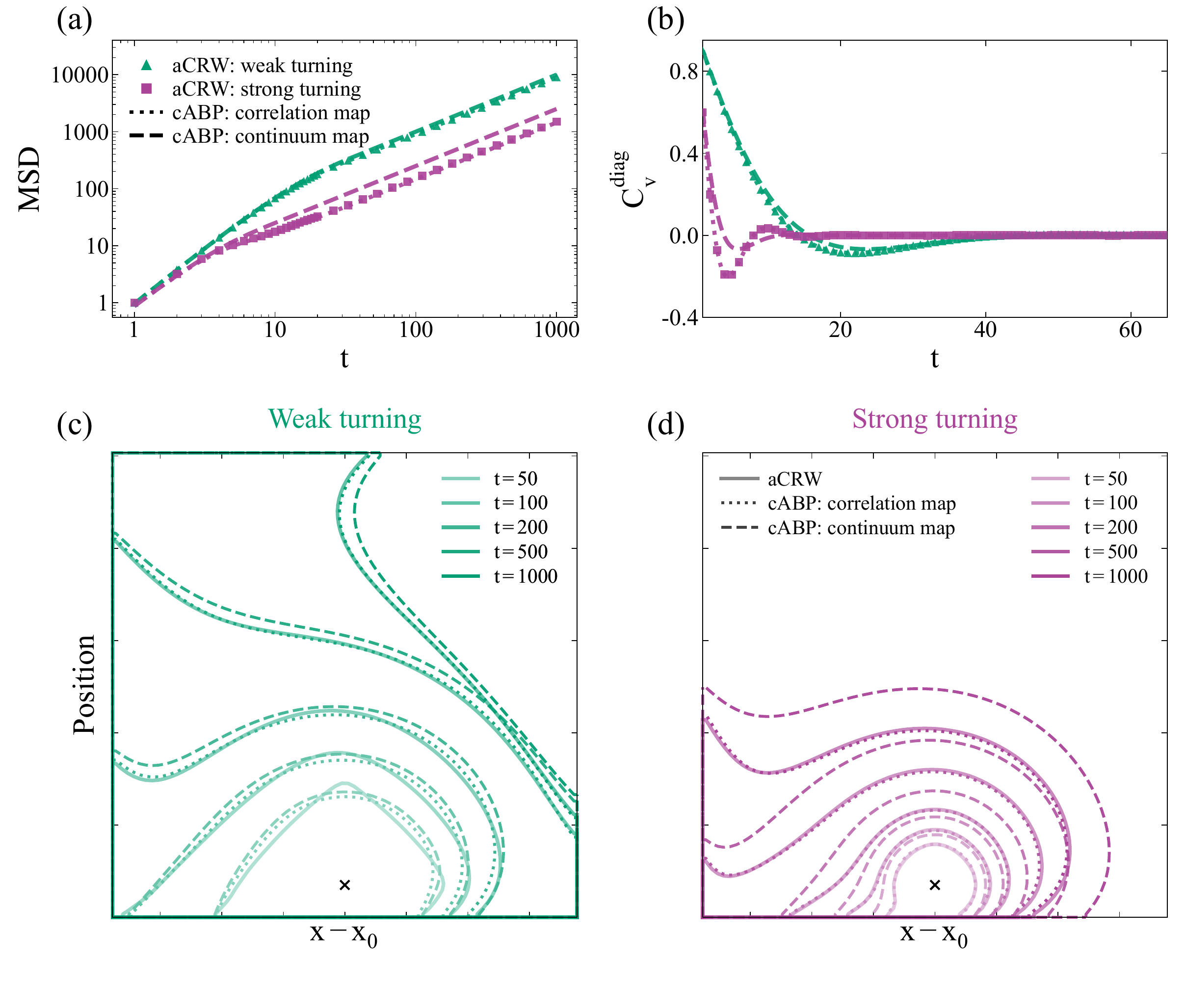}
    \caption{Comparison of the aCRW and cABP in the weak- and strong-turning regimes, with $\tilde D=0$ in all cABP calculations.
Green denotes weak turning, with $q_p=0.9$ and $q_c=0.1$, whereas purple denotes strong turning, with $q_p=0.6$ and $q_c=0.4$.
We use lattice units $\ell=\Delta t=1$.
(a) MSD as a function of time.
(b) Diagonal velocity autocorrelation $C_v^{\mathrm{diag}}$.
Markers show aCRW ensemble averages over $N=10^4$ trajectories.
(c,d) Contours enclosing $90\%$ of the probability in the highest-density regions for weak and strong turning, respectively.
Both models start with uniform orientations and the same Gaussian spatial distribution, centred at $(x_0,y_0)=(75.5,10.5)$ in a square domain of side length $151$.
The Gaussian has standard deviation $\sigma\approx2.043$ per coordinate, corresponding to a nominal radius $R_{95}=5$ containing $95\%$ of its unbounded probability, and is conditioned to lie inside the domain.
Crosses mark its centre.
All four walls are specularly reflecting, see Appendix \ref{app:boundary-scattering} for details.
Solid contours show the aCRW probability density obtained by iterating the lattice master equation.
The cABP densities are obtained numerically using a finite-volume spatial grid with spacing of $0.5$, $64$ angular bins, and integration time step $0.2$.
Both densities are binned onto a common $76\times76$ spatial grid before contour extraction.
Lighter to darker shades indicate $t=50,100,200,500,1000$.
Throughout the figure, dashed curves and contours use the continuum mapping in Eq.~\eqref{eq:relating coefficients}, whereas dotted curves and contours use the correlation map in Eq.~\eqref{eq:exact-correlation-mapping}.
The continuum mapping captures the asymptotic transport correspondence in the weak-turning limit, whereas the correlation map reproduces the bulk first angular harmonic exactly at integer update times.
For the two parameter sets shown, the correlation-matched cABP also agrees well with the measured bulk MSD, even though correlation matching alone does not guarantee agreement of the MSD or the confined spatial distributions.}
    \label{fig:VACF}
\end{figure}

Figure~\ref{fig:VACF} illustrates these distinctions.
The leading map describes the asymptotic transport correspondence, while the correlation map reproduces the sampled bulk first angular harmonic exactly.
For both parameter sets shown, the correlation-matched cABP also agrees well with the measured MSD, although correlation matching alone does not guarantee this agreement.
The relative long-time diffusion errors are $0.2\%$ and $3.7\%$ for correlation matching, versus $11\%$ and $67\%$ for the leading map (weak and strong turning, respectively).
Neither map identifies the higher angular harmonics or the full spatial distributions; the confined contours additionally depend on propagation and wall scattering.

\section*{Conclusion}
\label{sec:conclusions}
We introduced a minimal, intuitive, and computationally efficient active chiral random walk in which persistence, reversal, chirality, and randomization are controlled directly by a small set of microscopic probabilities with explicit dynamical meanings. The complex polar eigenvalue of the coin operator, \(\lambda=q_p+\mathrm{i}q_c\), governs the exact directional correlations, finite-time MSD, and leading second-order bulk transport coefficients, while $q_n$ controls nematic and higher-gradient information outside the regime analysed here.
The hydrodynamic and discrete Green--Kubo calculations yield identical long-time even and odd coefficients, showing which transport information survives the coarse-graining.

In the weak-turning limit generated by infrequent finite-angle reorientations, the aCRW reproduces the first-angular-harmonic dynamics and leading transport of the cABP. Beyond this correspondence, the minimal model provides a transparent benchmark for a general issue in coarse-graining: a retained mode can have approximate transient dynamics while the leading transport remains exact. Here both the omitted temporal corrections and the preserved lattice coefficients can be traced directly to the microscopic probabilities. Matching sampled correlations also leaves a distinction between their discrete sum and the continuous-time integral governing cABP transport. The four-state model does not become a continuous angular process: strong turning, deterministic chiral cycles, and anti-persistent motion remain intrinsically discrete regimes.

A distinctive result is the closed density-current relation
\[
\mathbf j=-D_{\mathrm{eff}}\nabla\rho-D_{\mathrm{odd}}\boldsymbol{\epsilon}\nabla\rho,
\]
in which chirality generates an odd current perpendicular to the density gradient. This term is divergence-free in a homogeneous bulk. At a boundary, however, rejected bonds and the microscopic scattering kernel alter the outgoing orientational populations, so wall accumulation must be interpreted jointly with the boundary rule rather than from $D_{\rm odd}$ alone.

The model is especially suited to confined, disordered, networked, and interacting systems because its microscopic rules remain transparent and inexpensive to simulate. 
It also suggests a direct experimental realization with programmable self-propelled robots or ``Hexbugs'', for which persistence, reversal, and handed turning can be imposed as discrete control instructions and both trajectories and boundary currents can be measured~\cite{kiechl2026free, dauchot2019dynamics}.
Further extensions could test how interactions, external forcing, or spatially varying parameters convert the transverse probability current into measurable odd mobility or Hall-like response. 
Beyond the MSD, the model also provides a tractable setting for studying non-Gaussian displacement statistics in active motion~\cite{khadem2021transport,lemaitre2023non}. Its finite microscopic propagation speed may likewise be useful in search problems for which telegrapher-like dynamics alter extreme first-passage statistics~\cite{grebenkov2026fastest,lawley2026universality}. The exact directional correlations also make the velocity power spectrum directly accessible, providing an experimentally convenient frequency-domain characterization of persistence and chirality.

\section*{Data accessibility}

The Python source code and figure-generation scripts supporting this article are archived as \emph{aCRW codes}, version v1.2.0~\cite{sherf2026aCRWcodes-v120}: \url{https://doi.org/10.5281/zenodo.24060170}. No external experimental data were used in this work.

\section*{Acknowledgements}

N. S. and E. K. acknowledge support from the Tel Aviv University--University of 
Potsdam "Research Cooperation Program" financing mutual 3-month stays that 
established their collaboration and eventually led to this work. R. M. 
acknowledges support by the Deutsche Forschungsgemeinschaft (grants No. ME 
1535/22-1, and No. 318763901 within the CRC 1294 ``Data Assimilation'', 
project B10). Y.R. and  N.S. acknowledge support from the Israel Science
Foundation (grants No. 385/21 and 2013/25).

\appendix
\renewcommand{\theequation}{\thesection\arabic{equation}}

\section{Derivation of the directional correlations}
\label{app:correlations}

Let \(d_t\in D\) denote the internal orientation at time \(t\), and let
\(\mathbf e_{d_t}=\mathrm E\ket{d_t}\) be the corresponding direction in the two-dimensional physical space, with \(\mathrm E\in\mathbb R^{2\times4}\) defined in Eq.~\eqref{eq:geometry-matrix}. 
The following derivation applies to the spatially homogeneous walk, for which the same coin \(\mathrm C\) is used at every update. 
Although \(\mathrm C\) does not determine the direction of an individual stochastic realization, it determines its conditional mean. 
At a lag of \(n\) updates each occurring after \(\Delta t\),
\begin{equation}
    \mathbb E\!\left[
        \mathbf e_{d_{t+n\Delta t}}
        \mid d_t=d
    \right]
    =
     \mathrm E \mathrm C^{\,n}\ket d
    =
     \mathrm C_{(2)}^{\,n}\mathrm E\ket d
    =
     \mathrm C_{(2)}^{\,n}\mathbf e_d,
    \label{eq:conditional-direction-n}
\end{equation}
where \(\mathrm E \mathrm C=\mathrm C_{(2)}\mathrm E\), with \(\mathrm C_{(2)}=q_p\mathbf 1+q_c\boldsymbol\epsilon\) as introduced in Eq.~\eqref{eq:polar-coin-main}. Here \(\mathbf 1\in\mathbb R^{2\times2}\) is the identity matrix and \(\boldsymbol\epsilon\in\mathbb R^{2\times2}\) generates a counter-clockwise rotation by \(\pi/2\).

For an isotropic stationary distribution, each lattice direction occurs with probability \(1/4\), and hence its expectation is
\begin{equation}
    \left\langle
        \mathbf e_{d_t} \otimes \mathbf e_{d_t}
    \right\rangle
    =
    \frac{1}{4}
    \sum_{d\in D}
        \mathbf e_d\mathbf e_d^{\mathrm T}
    =
    \frac{1}{2}\mathbf 1.
    \label{eq:isotropic-direction-second-moment}
\end{equation}
Conditional averaging of Eq.~\eqref{eq:conditional-direction-n} then gives the normalised velocity-autocorrelation tensor (VACT) used in the main text,
\begin{equation}
    \mathbf C_v(n)
    \equiv
    2\left\langle
        \mathbf e_{d_{t+n\Delta t}}\otimes\mathbf e_{d_t}
    \right\rangle
    =
    2\mathrm C_{(2)}^{\,n}
    \left\langle
        \mathbf e_{d_t}\otimes\mathbf e_{d_t}
    \right\rangle
    =\mathrm C_{(2)}^{\,n}.
    \label{eq:direction-correlation-reduced-coin}
\end{equation}
In particular, \(\mathbf C_v(0)=\mathbf 1\). Its decomposition into diagonal and antisymmetric parts is~\cite{hargus2021odd,kalz2024oscillatory}
\begin{equation}
    \mathbf C_v(n)
    =C_v^{\rm diag}(n)\mathbf 1
    +C_v^{\rm off}(n)\boldsymbol\epsilon,
    \label{eq:appendix-correlation-decomposition}
\end{equation}
where
\begin{align}
    C_v^{\rm diag}(n)
    &=\left\langle
        \mathbf e_{d_{t+n\Delta t}}\mathbin{\cdot}\mathbf e_{d_t}
      \right\rangle,
    \label{eq:longitudinal-correlation}\\
    C_v^{\rm off}(n)
    &=\left\langle
        \mathbf e_{d_{t+n\Delta t}}\mathbin{\cdot}
        \boldsymbol\epsilon\mathbf e_{d_t}
      \right\rangle.
    \label{eq:transverse-correlation}
\end{align}

The eigenvalues of \(\mathrm C_{(2)}\) are the conjugate polar pair \(\lambda=q_p+\mathrm i q_c\) and \(\overline\lambda\), already obtained from the full coin in Eqs.~\eqref{eq:lambda1} and \eqref{eq:lambda3}. Writing \(\lambda=|\lambda|\exp(\mathrm i\Omega)\), with \(|\lambda|=\sqrt{q_p^2+q_c^2}\) and \(\Omega=\arg\lambda\), gives
\begin{equation}
    \mathrm C_{(2)}^{\,n}
    =
    |\lambda|^n
    \left[
        \cos(n\Omega)\,\mathbf 1
        +
        \sin(n\Omega)\,\boldsymbol\epsilon
    \right]
    =
    |\lambda|^n
    \begin{pmatrix}
        \cos(n\Omega) & -\sin(n\Omega)\\
        \sin(n\Omega) & \cos(n\Omega)
    \end{pmatrix},
    \label{eq:polar-coin-n}
\end{equation}
The sign of the off-diagonal entries follows from the counter-clockwise-positive convention for \(\boldsymbol\epsilon\). Comparing Eqs.~\eqref{eq:appendix-correlation-decomposition} and \eqref{eq:polar-coin-n} yields
\begin{equation}
    C_v^{\rm diag}(n)=|\lambda|^n\cos(n\Omega)
    =\operatorname{Re}(\lambda^n),
    \qquad
    C_v^{\rm off}(n)=|\lambda|^n\sin(n\Omega)
    =\operatorname{Im}(\lambda^n),
\end{equation}
which are Eqs.~\eqref{eq:VACF} and \eqref{eq:VCCF}. Thus \(|\lambda|\) sets the decay of polar memory and \(\Omega\) sets its rotation and handedness. The nematic eigenvalue \(q_n\) does not enter either correlation.

\section{Polar reduction and hydrodynamic expansion}
\label{app:polar-reduction}

\subsection{Projection onto the orientation basis}

In the ordered orientation basis \(D=\{\rightarrow,\uparrow,\leftarrow,\downarrow\}\), the operator \(\boldsymbol\epsilon\) introduced in Eq.~\eqref{eq:polar-coin-main} rotates physical vectors counter-clockwise by \(\pi/2\). Thus a clockwise turn, reversal, and counter-clockwise turn are represented in the physical plane by \(\boldsymbol\epsilon^3\), \(\boldsymbol\epsilon^2\), and \(\boldsymbol\epsilon\), respectively. Projecting the reduced coin with the geometry matrix gives
\begin{equation}
    \mathrm E \mathrm C=
    \left(
        \hat q_I\,\mathbf 1
        +\hat q_R\,\boldsymbol\epsilon^3
        +\hat q_B\,\boldsymbol\epsilon^2
        +\hat q_L\,\boldsymbol\epsilon
    \right)\mathrm E,
\label{eq:coin-projection-expanded}
\end{equation}
where, in two-dimensional space,
\begin{equation}
    \boldsymbol\epsilon^2=-\mathbf 1,
    \qquad
    \boldsymbol\epsilon^3=-\boldsymbol\epsilon,
    \qquad
    \boldsymbol\epsilon^4=\mathbf 1.
    \label{eq:levi-civita-powers}
\end{equation}
It follows that
\begin{equation}
    \mathrm E \mathrm C =
    \left[
        (\hat q_I-\hat q_B)\mathbf 1
        +
        (\hat q_L-\hat q_R)\boldsymbol\epsilon
    \right]\mathrm E =
     \left(q_p\,\mathbf 1+q_c\,\boldsymbol\epsilon\right)\mathrm E
    \equiv
    \mathrm C_{(2)} \mathrm E.
    \label{eq:coin-intertwining}
\end{equation}
The polar projection therefore retains only the two contrasts \(q_p = \hat q_I - \hat q_B\) and \(q_c = \hat q_L - \hat q_R\). Since the eigenvalues of \(\boldsymbol\epsilon\) are \(\pm\mathrm i\), the eigenvalues of \(\mathrm C_{(2)}\) are \(q_p\pm\mathrm i q_c=\lambda_{1,3}\), as in Eqs.~\eqref{eq:lambda1} and \eqref{eq:lambda3}.

\subsection{Gradient expansion and moment equations}
\label{app:continuum-derivation}

We now show explicitly where the omitted nematic mode enters the gradient hierarchy. Let \(\mathbf P(\mathbf r,t)=(P_{\rightarrow},P_{\uparrow},P_{\leftarrow},P_{\downarrow})^{\mathrm T}(\mathbf r,t)\). Its scalar, polar, and nematic moments are
\begin{equation}
    \rho(\mathbf r,t)=\mathbf 1_4^{\mathrm T}\mathbf P(\mathbf r,t),
    \qquad
    \mathbf m(\mathbf r,t)=\mathrm E\mathbf P(\mathbf r,t),
    \qquad
    a(\mathbf r,t)=P_{\rightarrow}(\mathbf r,t)+P_{\leftarrow}(\mathbf r,t)
      -P_{\uparrow}(\mathbf r,t)-P_{\downarrow}(\mathbf r,t),
    \label{eq:appendix-moments}
\end{equation}
where \(\mathbf 1_4=(1,1,1,1)^{\mathrm T}\), and the last quantity is the amplitude of the symmetric traceless nematic tensor
\begin{equation}
    \mathbf N(\mathbf r,t)
    =\frac{a(\mathbf r,t)}{2}
      \begin{pmatrix}1&0\\0&-1\end{pmatrix}.
\end{equation}
Greek indices below denote Cartesian coordinates and repeated Greek indices are summed. Expanding the master equation, Eq.~\eqref{eq:master-equation}, for fields that vary slowly over one lattice spacing gives
\begin{equation}
    \frac{\partial P_d(\mathbf r,t)}{\partial t}
    =
    \nu
    \left\{
    \sum_{d'}
    C_{dd'}
    \left[
        1-\ell e_{d,\alpha}\nabla_\alpha
        +\frac{\ell^2}{2}e_{d,\alpha}e_{d,\beta}
          \nabla_\alpha\nabla_\beta
    \right]P_{d'}(\mathbf r,t)
    -P_d(\mathbf r,t)
    \right\}
    +\mathcal O(\ell^3\nabla^3).
    \label{eq:appendix-expanded-master}
\end{equation}

The first outgoing moment is \(\mathrm C_{(2)}\mathbf m\) by Eq.~\eqref{eq:coin-intertwining}. The second outgoing moment is
\begin{equation}
    M^{\rm out}_{\alpha\beta}(\mathbf r,t)
    \equiv\sum_{d,d'}e_{d,\alpha}e_{d,\beta}C_{dd'}P_{d'}(\mathbf r,t),
    \qquad
    \mathbf M^{\rm out}
    =
    \frac{1}{2}
    \begin{pmatrix}
        \rho(\mathbf r,t)+q_n a(\mathbf r,t)&0\\
        0&\rho(\mathbf r,t)-q_n a(\mathbf r,t)
    \end{pmatrix},
    \label{eq:appendix-second-moment}
\end{equation}
where \(q_n\) is the nematic eigenvalue in Eq.~\eqref{eq:lambda2}. Summing Eq.~\eqref{eq:appendix-expanded-master} over \(d\) now produces
\begin{equation}
    \frac{\partial \rho(\mathbf r,t)}{\partial t}
    =
    -\nu\ell
        \nabla\mathbin{\cdot}
        \left[\mathrm C_{(2)}\mathbf m(\mathbf r,t)\right]
    +\frac{\nu\ell^2}{4}
        \left[
            \nabla^2\rho(\mathbf r,t)
            +q_n(\nabla_x^2-\nabla_y^2)a(\mathbf r,t)
        \right]
    +\mathcal O(\ell^3\nabla^3\mathbf m).
    \label{eq:appendix-density-expansion}
\end{equation}
The operator \(\nabla_x^2-\nabla_y^2\) here is the first explicit imprint of the Cartesian lattice symmetry.

Projecting the same expanded master equation onto the polarization yields
\begin{equation}
    \frac{\partial \mathbf m(\mathbf r,t)}{\partial t}
    =
    -\nu
    \left[
        (1-q_p)\mathbf 1
        -q_c\boldsymbol\epsilon
    \right]\mathbf m(\mathbf r,t)
    -\frac{\nu\ell}{2}\nabla\rho(\mathbf r,t)
    -\frac{\nu\ell q_n}{2}
      \begin{pmatrix}\nabla_x a(\mathbf r,t)\\-\nabla_y a(\mathbf r,t)\end{pmatrix}
    +\mathcal O(\ell^2\nabla^2\mathbf m).
    \label{eq:appendix-polarization-dynamics}
\end{equation}
Finally, projecting onto \(a\) gives, to the order needed here,
\begin{align}
    \frac{\partial a(\mathbf r,t)}{\partial t}
    ={}&-\nu(1-q_n)a(\mathbf r,t)
    -\nu\ell\left[
        \nabla_x(\mathrm C_{(2)}\mathbf m(\mathbf r,t))_x
        -\nabla_y(\mathrm C_{(2)}\mathbf m(\mathbf r,t))_y
      \right]
    \notag\\
    &+\frac{\nu\ell^2}{4}
      \left[(\nabla_x^2-\nabla_y^2)\rho(\mathbf r,t)
      +q_n\nabla^2a(\mathbf r,t)\right]
    +\mathcal O(\ell^3\nabla^3\mathbf m).
    \label{eq:appendix-nematic-dynamics}
\end{align}
Equations~\eqref{eq:appendix-polarization-dynamics} and \eqref{eq:appendix-nematic-dynamics} make the closure transparent. Provided the nematic mode is non-conserved, \(1-q_n\neq0\), the density first generates \(\mathbf m=\mathcal O(\nabla\rho)\), while \(a\) is generated only at \(\mathcal O(\nabla^2\rho)\). Its feedback into the polarization is therefore third order in gradients, and its explicit contribution to Eq.~\eqref{eq:appendix-density-expansion} is fourth order. The second-order bulk theory consequently retains only \(\rho\) and \(\mathbf m\); keeping \(q_n\) consistently would also require the square-lattice, fourth-gradient corrections. The quasi-stationary elimination and the resulting even and odd currents are carried out in the main text, from Eq.~\eqref{eq:polarization-eq} through Eq.~\eqref{eq:effective_diff_coeffs}.

\subsection{Microscopic probability current}
\label{app:microscopic-current}

The microscopic current is fixed by the probability transported along lattice bonds.
Since the coin acts before the displacement, the outgoing populations are
\begin{equation}
    P_d^{+}(\mathbf r,t)
    =
    \sum_{d'}C_{dd'}P_{d'}(\mathbf r,t).
    \label{eq:post-coin-population}
\end{equation}
The one-way probability flux from \(\mathbf r\) to \(\mathbf r+\ell\mathbf e_d\) is
\begin{equation}
    j_d\!\left(\mathbf r+\frac{\ell}{2}\mathbf e_d,t\right)
    =
    \nu P_d^{+}(\mathbf r,t).
    \label{eq:directed-bond-current}
\end{equation}
The argument locates this flux at the bond midpoint; the displacement remains the full \(\ell\mathbf e_d\).
Weighting the outgoing fluxes at a site by their displacements gives
\begin{equation}
    \mathbf j_{\rm lat}(\mathbf r,t)
    =
    \nu\ell\sum_d\mathbf e_d P_d^{+}(\mathbf r,t)
    =
    \nu\ell\,\mathrm E\mathrm C\mathbf P(\mathbf r,t)
    =
    \nu\ell\,\mathrm C_{(2)}\mathbf m(\mathbf r,t),
    \label{eq:appendix-outgoing-current}
\end{equation}
where Eq.~\eqref{eq:coin-intertwining} gives the final equality.
This establishes Eq.~\eqref{eq:lattice-vector-current} and distinguishes the current from the pre-coin polarization.

To express the bond fluxes at a common spatial point, their source populations are evaluated at \(\mathbf r-\ell\mathbf e_d/2\).
Expanding these populations to first order in spatial gradients leads to
\begin{equation}
    j_\alpha(\mathbf r,t)
    \approx
    (j_{\rm lat})_\alpha(\mathbf r,t)
    -
    \frac{\nu\ell^2}{2}\nabla_\beta
    M^{\rm out}_{\alpha\beta}(\mathbf r,t),
    \label{eq:appendix-bond-current-expansion}
\end{equation}
with the outgoing second moment defined in Eq.~\eqref{eq:appendix-second-moment}.
In the second-order bulk closure, its leading part is
\(M^{\rm out}_{\alpha\beta}\approx\rho\,\delta_{\alpha\beta}/2\), so
\begin{equation}
    \mathbf j(\mathbf r,t)
    \approx
    \mathbf j_{\rm lat}(\mathbf r,t)-D_0\nabla\rho(\mathbf r,t),
    \qquad
    D_0=\frac{\nu\ell^2}{4}.
    \label{eq:appendix-continuum-current}
\end{equation}
The omitted nematic and higher-gradient contributions are beyond the retained bulk order.
Substituting the quasi-stationary polarization, Eq.~\eqref{eq:m-closure}, recovers Eq.~\eqref{eq:current-even-odd} with the coefficients in Eq.~\eqref{eq:effective_diff_coeffs}.
Thus the odd coefficient belongs to a specified microscopic current, even though its homogeneous contribution disappears from the scalar density equation.

\section{Boundary scattering rules}
\label{app:boundary-scattering}

\begin{figure}
    \centering
    \includegraphics[width=0.9\linewidth]{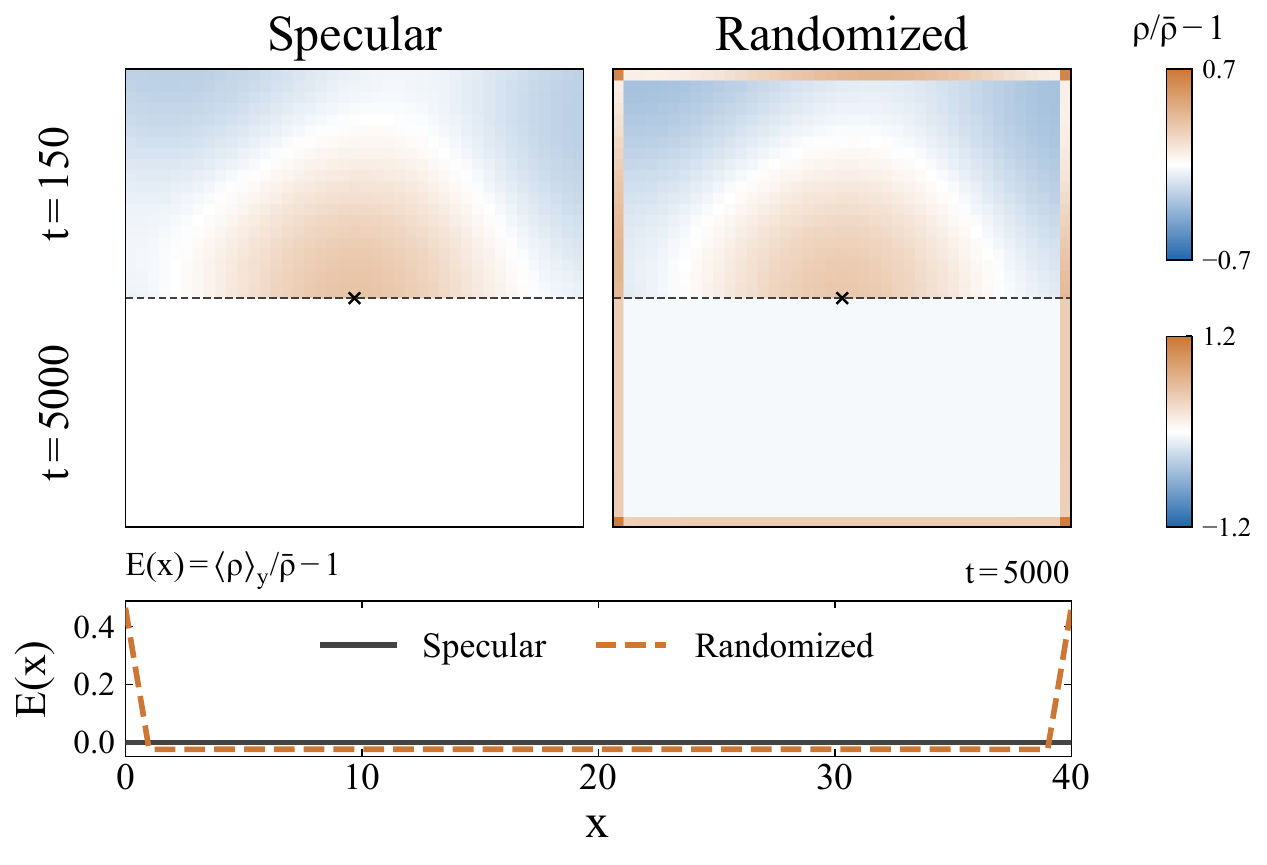}
    \caption{Boundary-dependent density evolution for $q_p=0.6$ and $q_c=0.2$ on a $41\times41$ lattice, starting from a Gaussian distribution restricted to the central $5\times5$ sites with uniform orientations.
    Each square combines the upper half-domain at $t=150$ and the lower half-domain at $t=5000$; the dashed line separates the two times.
    Colours show the density excess $\rho/\bar\rho-1$, with separate scales for each time.
    Specular reflection produces a uniform stationary density, whereas randomized scattering produces wall accumulation.
    The lower panel shows the final projected density excess $E(x)=\langle\rho\rangle_y/\bar\rho-1$.
    Densities are obtained by deterministic iteration of the lattice master equation, with $\ell=\Delta t=1$.}
    \label{fig:boundary-rules}
\end{figure}

In the homogeneous bulk, one update is generated by
\begin{equation}
    \mathrm U_{\rm bulk}
    =\mathrm S(\mathrm I_{\mathbf r}\otimes\mathrm C),
\end{equation}
where \(\mathrm C\) updates the orientation and \(\mathrm S\) then shifts the walker in that direction, as in Eq.~\eqref{eq:evolution}.
Confinement requires an additional prescription for attempted displacements that cross a wall.
We adopt specular reflection as the reference boundary rule as it is a standard idealized reflecting condition used in active-particle models and can be applied consistently to both the aCRW and cABP~\cite{escalante2026benchmarking, granek2024colloquium}.
Randomized scattering is considered in Fig.~\ref{fig:boundary-rules} to illustrate the dependence on this microscopic choice.

At each update, the bulk coin first selects a proposed outgoing orientation \(d\).
If the target site lies inside the domain, the displacement is accepted.
Otherwise, the displacement is rejected, the position remains unchanged, and the wall produces a new orientation \(d'\).
Time advances by one step in either case.
The scattering is specified by the conditional probability
\begin{equation}
    K_{\mathrm{wall}}(d'\mid d)
    ,
    \qquad d'\in D.
    \label{eq:wall-kernel}
\end{equation}

For specular reflection at a wall with outward unit normal \(\mathbf n\), the normal component of the heading is reversed while its tangential component is preserved. In continuous space this is 
\begin{equation}
    \mathbf e_{d'}
    =
    \mathbf e_d
    -2(\mathbf e_d\cdot\mathbf n)\mathbf n.
    \label{eq:wall-specular}
\end{equation}
With four cardinal headings and axis-aligned walls, a rejected displacement is normal to the wall.
Specular reflection therefore reduces to reversal of the orientation,
\(K_{\mathrm{wall}}^{\mathrm{spec}}(d'\mid d)
=\delta_{d',\mathrm B d}\).
This is the boundary rule used for the confined calculations in the main text.

An alternative is randomized scattering,
\begin{equation}
    K_{\mathrm{wall}}^{\mathrm{rand}}(d'\mid d)
    =\frac14
    \label{eq:wall-random}
\end{equation}
which redistributes the rejected probability equally among all four headings, including those pointing outside.
Subsequent attempted displacements remain subject to the same rejection rule.

Figure~\ref{fig:boundary-rules} compares the two prescriptions at identical bulk parameters.
For the parameters shown, specular reflection yields a uniform stationary density, whereas randomized scattering produces enhanced wall occupation.
The boundary dynamics therefore provide an additional microscopic ingredient: wall accumulation depends on the scattering prescription and is not determined by the bulk odd-diffusion coefficient alone~\cite{granek2024colloquium,escalante2026benchmarking}.

\section{Derivation of the cABP field equation}
\label{app:cABP-marginalization}

For completeness, we derive the cABP density and polarization equations used in Sec.~\ref{sec:cABP-connection}. In continuous two-dimensional space, the position \(\mathbf r(t)\) and orientation angle \(\phi(t)\) obey
\begin{equation}
    d\mathbf r=\tilde v_0\mathbf u(\phi)\,dt
    +\sqrt{2\tilde D}\,d\mathbf B,
    \qquad
    d\phi=\tilde\omega\,dt+\sqrt{2\tilde D_r}\,dW,
\end{equation}
where \(\mathbf u(\phi)=(\cos\phi,\sin\phi)\), $d\mathbf B$ and $dW$ are independent Wiener increments with unit variance. The corresponding probability density \(\tilde p(\mathbf r,\phi,t)\) satisfies
\begin{equation}
    \frac{\partial \tilde p(\mathbf r,\phi,t)}{\partial t}
    =\tilde D\nabla^2\tilde p
    -\tilde v_0\mathbf u\mathbin{\cdot}\nabla \tilde p
    +\tilde D_r\frac{\partial^2\tilde p}{\partial\phi^2}
    -\tilde\omega\frac{\partial \tilde p}{\partial\phi}.
    \label{eq:appendix-cabp-fokker-planck}
\end{equation}
This is Eq.~\eqref{eq:cabp-fokker-planck}, written here to make the marginalization self-contained. We use the counter-clockwise-positive convention, so \(\tilde\omega>0\) increases \(\phi\). The angular identities required below are
\begin{equation}
    \frac{\partial\mathbf u}{\partial\phi}
    =\boldsymbol{\epsilon}\mathbf u,
    \qquad
    \frac{\partial^2\mathbf u}{\partial\phi^2}
    =-\mathbf u.
    \label{eq:cabp-angular-derivatives}
\end{equation}

\subsection{Angular marginalization}

The first two angular moments are
\begin{equation}
    \tilde\rho(\mathbf r,t)=\int_0^{2\pi}\tilde p(\mathbf r,\phi,t)\,d\phi,
    \qquad
    \tilde{\mathbf m}(\mathbf r,t)
    =\int_0^{2\pi}\mathbf u(\phi)\tilde p(\mathbf r,\phi,t)\,d\phi.
\end{equation}
Integrating Eq.~\eqref{eq:appendix-cabp-fokker-planck} over \(\phi\), periodicity eliminates both angular-derivative terms and \(\int_0^{2\pi}\mathbf u\mathbin{\cdot}\nabla \tilde p\,d\phi=\nabla\mathbin{\cdot}\tilde{\mathbf m}\). Hence
\begin{equation}
    \frac{\partial\tilde\rho(\mathbf r,t)}{\partial t}
    =\tilde D\nabla^2\tilde\rho(\mathbf r,t)
    -\tilde v_0\nabla\mathbin{\cdot}\tilde{\mathbf m}(\mathbf r,t),
\end{equation}
which is Eq.~\eqref{eq:cabp-density}.

To obtain the first-moment equation, we multiply Eq.~\eqref{eq:appendix-cabp-fokker-planck} by \(\mathbf u(\phi)\) and integrate over \(\phi\). Integration by parts and Eq.~\eqref{eq:cabp-angular-derivatives} give
\begin{equation}
\int_0^{2\pi}\mathbf u\,\frac{\partial \tilde p}{\partial\phi}\,d\phi
=-\boldsymbol\epsilon\tilde{\mathbf m},
\qquad
\int_0^{2\pi}\mathbf u\,\frac{\partial^2\tilde p}{\partial\phi^2}\,d\phi
=-\tilde{\mathbf m}.
\end{equation}
The propulsion term contains the second angular moment
\begin{equation}
M_{\alpha\beta}(\mathbf r,t)
=\int_0^{2\pi}u_\alpha u_\beta \tilde p(\mathbf r,\phi,t)\,d\phi
=\frac{\tilde\rho(\mathbf r,t)}{2}\delta_{\alpha\beta}
+\tilde N_{\alpha\beta}(\mathbf r,t),
\end{equation}
where Greek indices denote Cartesian components and \( \tilde {\mathbf{N}}\) is the symmetric traceless nematic tensor. The exact first-moment equation is therefore
\begin{align}
    \frac{\partial\tilde{\mathbf m}(\mathbf r,t)}{\partial t}
    ={}&\tilde D\nabla^2\tilde{\mathbf m}(\mathbf r,t)
    -\tilde D_r\tilde{\mathbf m}(\mathbf r,t)
    +\tilde\omega\boldsymbol\epsilon\tilde{\mathbf m}(\mathbf r,t)
    -\frac{\tilde v_0}{2}\nabla\tilde\rho(\mathbf r,t)
    -\tilde v_0\nabla\mathbin{\cdot}\tilde {\mathbf{N}}(\mathbf r,t).
    \label{eq:appendix-cabp-exact-polarization}
\end{align}
Here \((\nabla\mathbin{\cdot}\tilde {\mathbf{N}})_\alpha=\nabla_\beta \tilde N_{\alpha\beta}\). The first-harmonic, long-wavelength closure sets \(\tilde {\mathbf{N}}\approx0\), reducing Eq.~\eqref{eq:appendix-cabp-exact-polarization} to Eq.~\eqref{eq:cabp-polarization}. The sign of the angular-drift term is now explicit: positive \(\tilde\omega\) rotates the polarization counter-clockwise through \(+\tilde\omega\boldsymbol\epsilon\tilde{\mathbf m}\).

\subsection{Quasi-stationary closure}

In the slow-density, fast-polarization regime, neglecting \(\partial_t\tilde{\mathbf m}\) and \(\tilde D\nabla^2\tilde{\mathbf m}\) in the closed first-harmonic equation gives
\begin{align}
    \left(
        \tilde D_r\mathbf 1
        -
        \tilde\omega\boldsymbol{\epsilon}
    \right)\tilde{\mathbf m}(\mathbf r,t)
    &\approx
    -\frac{\tilde v_0}{2}\nabla\tilde\rho(\mathbf r,t),
    \notag\\
    \tilde{\mathbf m}(\mathbf r,t)
    &\approx
    -
    \frac{\tilde v_0}{2}
    \frac{
        \tilde D_r\mathbf 1
        +
        \tilde\omega\boldsymbol{\epsilon}
    }{
        \tilde D_r^2+\tilde\omega^2
    }
    \nabla\tilde\rho(\mathbf r,t).
\end{align}
The inverse follows from \(\boldsymbol\epsilon^2=-\mathbf 1\):
\begin{equation}
    \left(\tilde D_r\mathbf 1-\tilde\omega\boldsymbol\epsilon\right)^{-1}
    =\frac{\tilde D_r\mathbf 1+\tilde\omega\boldsymbol\epsilon}
    {\tilde D_r^2+\tilde\omega^2}.
\end{equation}
Substitution into \(\tilde{\mathbf j}(\mathbf r,t)=-\tilde D\nabla\tilde\rho(\mathbf r,t)+\tilde v_0\tilde{\mathbf m}(\mathbf r,t)\) gives the even and odd coefficients in Eq.~\eqref{eq:cabp-diffusion-coefficients}. The term proportional to \(\tilde D_r\) is antiparallel to the density gradient, whereas the term proportional to \(\tilde\omega\) is rotated counter-clockwise by \(\pi/2\).

\section{Finite-step corrections to the cABP comparison}
\label{app:finite-step-matching}

We detail the rate and transport corrections discussed in Sec.~\ref{sec:cABP-connection}.
Throughout, $\tilde D=0$, $\tilde v_0=\ell/\Delta t$, and the correlations refer to an isotropic, homogeneous bulk.

\subsection{Expansion of the correlation map}
\label{app:finite-step-generator}

Without spatial gradients, the exact polarization update is $\mathbf m(t+\Delta t)=\mathrm C_{(2)}\mathbf m(t)$.
Replacing the left-hand side by $\mathbf m+\Delta t\,\partial_t\mathbf m$ gives the leading rates in Eq.~\eqref{eq:relating coefficients}.
The exact correlation rates instead follow from $\lambda=\exp[(-\tilde D_r^{\rm corr}+\mathrm i\tilde\omega^{\rm corr})\Delta t]$.

Using $\lambda=1+(-\tilde D_r+\mathrm i\tilde\omega)\Delta t$ and expanding its logarithm gives
\begin{equation}
    \frac{\log\lambda}{\Delta t}
    =-\tilde D_r+\mathrm i\tilde\omega
    -\frac{\Delta t}{2}(-\tilde D_r+\mathrm i\tilde\omega)^2
    +\mathcal O(\Delta t^2).
    \label{eq:appendix-generator-series}
\end{equation}
Separating real and imaginary parts yields
\begin{align}
    \tilde D_r^{\rm corr}
    &=\tilde D_r+\frac{\Delta t}{2}
      \left(\tilde D_r^2-\tilde\omega^2\right)
      +\mathcal O(\Delta t^2),
    \label{eq:decay-rate-corrections}\\
\intertext{and}
    \tilde\omega^{\rm corr}
    &=\tilde\omega+\Delta t\,\tilde D_r\tilde\omega
      +\mathcal O(\Delta t^2).
    \label{eq:rotation-rate-corrections}
\end{align}
These expansions hold as $\Delta t\to0$ with bounded leading rates.
They quantify the temporal corrections omitted by the continuum map, without changing the exact polar projection.
We choose the principal phase $\arg\lambda\in(-\pi,\pi]$, continuous near $\lambda=1$; frequencies differing by $2\pi/\Delta t$ cannot be distinguished at the update times.
The case $\lambda=0$ has no finite correlation-matched decay rate.

\subsection{Sums and integrals of correlations}
\label{app:finite-step-transport}

Under correlation matching, the complex combination $C_v^{\rm diag}(n)+\mathrm iC_v^{\rm off}(n)=\lambda^n$ equals the cABP exponential at every update time.
For $0<|\lambda|<1$, the corresponding discrete sum and continuous integral are
\begin{align}
    \Delta t\left[\frac12+\sum_{n=1}^{\infty}\lambda^n\right]
    &=\frac{\Delta t}{2}\frac{1+\lambda}{1-\lambda},
    \label{eq:discrete-correlation-sum}\\
\intertext{and}
    \int_0^\infty
      \exp[(-\tilde D_r^{\rm corr}+\mathrm i\tilde\omega^{\rm corr})t]\,dt
    &=\frac{1}{\tilde D_r^{\rm corr}-\mathrm i\tilde\omega^{\rm corr}}.
    \label{eq:continuous-correlation-integral}
\end{align}
Multiplying their real and imaginary parts by $\tilde v_0^2/2$ gives the respective even and odd coefficients.
The half weight at zero lag is the normalization in Eq.~\eqref{eq:GK-discrete}.

Expanding the sum at fixed correlation-matched rates yields
\begin{align}
    \Delta t\left[\frac12+\sum_{n=1}^{\infty}\lambda^n\right]
    ={}&\frac{1}{\tilde D_r^{\rm corr}-\mathrm i\tilde\omega^{\rm corr}}
+\frac{\Delta t^2}{12}
       \left(\tilde D_r^{\rm corr}-\mathrm i\tilde\omega^{\rm corr}\right)
       +\mathcal O(\Delta t^4).
    \label{eq:correlation-sum-expansion}
\end{align}
Thus, with the cABP coefficients evaluated using the correlation map,
\begin{align}
    D_{\rm eff}-\tilde D_{\rm eff}
    &=\frac{\tilde v_0^2\tilde D_r^{\rm corr}\Delta t^2}{24}
      +\mathcal O(\Delta t^4),
    \label{eq:correlation-map-even-error}\\
\intertext{and}
    D_{\rm odd}-\tilde D_{\rm odd}
    &=-\frac{\tilde v_0^2\tilde\omega^{\rm corr}\Delta t^2}{24}
      +\mathcal O(\Delta t^4).
    \label{eq:correlation-map-odd-error}
\end{align}
This is the correction between the discrete, half-weighted sum and the continuous integral.
The expansion requires $\Delta t\sqrt{(\tilde D_r^{\rm corr})^2+(\tilde\omega^{\rm corr})^2}\ll1$; Eqs.~\eqref{eq:discrete-correlation-sum} and \eqref{eq:continuous-correlation-integral} remain exact outside this regime.

For the leading map, direct substitution into Eq.~\eqref{eq:cabp-diffusion-coefficients} instead gives
\begin{equation}
    \tilde D_{\rm eff}-D_{\rm eff}
    =\frac{\ell^2}{4\Delta t}
      \frac{2(1-q_p)-(1-q_p^2-q_c^2)}
           {(1-q_p)^2+q_c^2}
    =D_0.
    \label{eq:leading-map-even-offset}
\end{equation}
The odd coefficients coincide by the same substitution, proving Eq.~\eqref{eq:leading-map-transport-offset}.
These finite-step differences between models do not alter the equality of the lattice hydrodynamic and discrete Green--Kubo coefficients.

\end{document}